%% file: ms_astroph.tex
\documentclass[12pt,manuscript]{aastex}

\newcommand{\ltsimeq}{\la}

\received{21 Dec 2006}
\revised{26 Mar 2007}
\accepted{4 Apr 2007}
\shortauthors{McQuinn et al.}
\shorttitle{M33 IR Variable Stars}

\sfcode`A=1000 \sfcode`B=1000 \sfcode`C=1000 \sfcode`D=1000
\sfcode`E=1000 \sfcode`F=1000 \sfcode`G=1000 \sfcode`H=1000
\sfcode`I=1000 \sfcode`J=1000 \sfcode`K=1000 \sfcode`L=1000
\sfcode`M=1000 \sfcode`N=1000 \sfcode`O=1000 \sfcode`P=1000
\sfcode`Q=1000 \sfcode`R=1000 \sfcode`S=1000 \sfcode`T=1000
\sfcode`U=1000 \sfcode`V=1000 \sfcode`W=1000 \sfcode`X=1000
\sfcode`Y=1000 \sfcode`Z=1000

\begin{document}
\title{The M33 Variable Star Population Revealed by {\it Spitzer}}
\author{K.~B.~W. McQuinn\altaffilmark{1}, 
Charles E. Woodward\altaffilmark{1}, 
S.~P. Willner\altaffilmark{2}, 
E.~F. Polomski\altaffilmark{1}, 
R.~D. Gehrz\altaffilmark{1},
Roberta M. Humphreys\altaffilmark{1}, 
Jacco~Th.~van Loon\altaffilmark{3}, 
M.~L.~N. Ashby\altaffilmark{2}, \\
K. Eicher\altaffilmark{1},
G.~G. Fazio\altaffilmark{2}}

\altaffiltext{1}{Department of Astronomy, School of Physics and
Astronomy, 116 Church Street, S.E., University of Minnesota,
Minneapolis, MN 55455, \ {\it kmcquinn@astro.umn.edu}} 
\altaffiltext{2}{Harvard-Smithsonian Center for Astrophysics, 60
Garden Street, Cambridge, MA 02138} 
\altaffiltext{3}{Lennard-Jones Laboratories, School of Physical and 
Geographical Sciences, Keele University, Staffordshire ST5~5BG, UK} 


\begin{abstract}
We analyze five epochs of {\it Spitzer Space 
Telescope}/Infrared Array Camera (IRAC) observations of the 
nearby spiral galaxy M33. Each epoch covered nearly a square 
degree at 3.6, 4.5, and 8.0~$\mu$m. 
The point source catalog from the full dataset contains 37,650 stars.
The stars have luminosities characteristic of the 
asymptotic giant branch and can be separated into oxygen-rich and 
carbon-rich populations by their $[3.6] - [4.5]$ colors. The $[3.6] - [8.0]$ 
colors indicate that over 80\% of the stars detected at 8.0~$\mu$m 
 have dust shells. Photometric 
comparison of epochs using conservative criteria yields a catalog of 
2,923 variable stars. These variables are 
most likely long-period variables amidst an evolved stellar population. 
At least one-third of the identified carbon stars are 
variable. 
\end{abstract} 


\section{Introduction \label{intro} }
The Triangulum galaxy, M33, is a late-type spiral galaxy and the third 
largest galaxy in the Local Group after the Andromeda and the Milky Way 
galaxies. M33 is well-suited for high spatial resolution studies of 
stellar populations in spiral galaxies because it is only 830 kpc 
distant (Distance modulus of 24.60, \citet{Wil90}) and has a 
favorable inclination angle of 52 degrees 
\citep{Cor00}. For these reasons M33 has been the subject of 
numerous such studies covering 
many wavelengths including the radio \citep{Eng03, Ros07}, near-infrared 
\citep[][Two Micron All Sky Survey -- 2MASS]{Skr06}, optical \citep{Mas06, 
Mac01, Hum80}, and X-ray \citep{Pie04} regimes. 

Resolved stellar population studies of M33 at infrared (IR) wavelengths are 
now possible using the Infrared Array Camera \citep[][IRAC]{Faz04} on the 
\textit{Spitzer Space Telescope} \citep{Wer04, Gehrz07}. Even with modest 
integration times, IRAC can detect a significant segment 
of the most luminous stellar population in M33.  \citet{Hum06}, 
for example, presented a detailed 
analysis of the early-type star Variable~A.  More numerous sources 
detectable with IRAC at 3.6 and 4.5~$\mu$m include luminous, 
cooler stars that have evolved off 
the main sequence, i.e., giant and supergiant stars, whose energy 
distributions peak in the near-IR. The point sources 
detected at the longest IRAC wavelength (8.0~$\mu$m) include a 
population which exhibits 
long-period, large-amplitude variability and produces extended 
circumstellar winds resulting in significant mass loss. The dust grain 
formation arising in these winds causes the spectral energy 
distributions of these stars to deviate from 
the Rayleigh-Jeans tail of a blackbody curve 
typical of a stellar photosphere, resulting in an IR excess at 
8.0~$\mu$m. These stars can be partially or completely obscured at optical 
wavelengths by the same dust whose thermal emission makes 
them stand out at IR wavelengths. \citet{Jac07} found that up to 39\% 
of the IRAC detected asymptotic giant branch (AGB) stars in the Local 
Group dwarf galaxy WLM are not detected at optical wavelengths. It is 
therefore plausible that a portion of the infrared-luminous M33 stellar 
populations are distinct from those previously cataloged at 
visible wavelengths and are only now beginning to be understood.

The knowledge of variable stars in other Local Group galaxies has
increased considerably in recent years with the advent of
large-scale, high-resolution mosaic imaging technology and
microlensing projects such as EROS \citep{Aub93} and MACHO
\citep{Alc97}, which have cataloged variable stars as a byproduct of
their surveys. However, few high resolution variable star studies
have encompassed the entire M33 galaxy because of its sheer angular
size. Two exceptions are the DIRECT optical project \citep{Mac01},
which imaged the center region of M33 ($R < 5$~kpc) cataloging
$\simeq$10$^3$ variable stars, and a recent optical CFHT survey by
\citet{Har06}, which imaged $\simeq$1~deg$^2$ creating a catalog of
over 10$^4$ variables.

\begin{figure}
\plotone{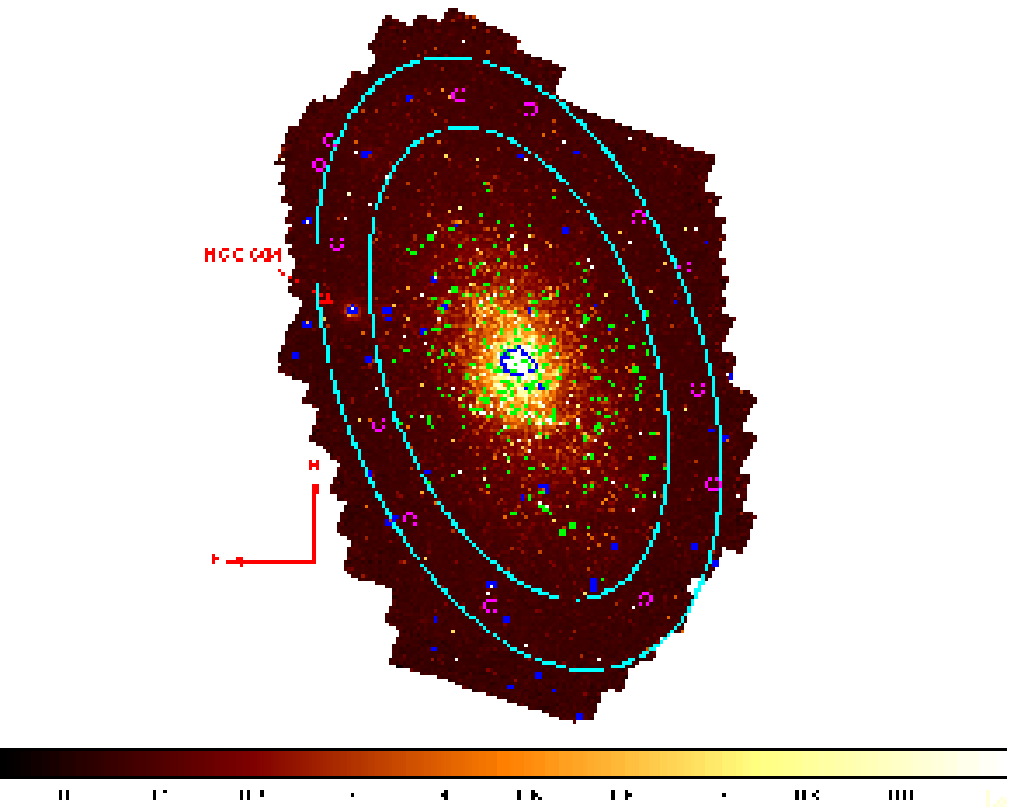}
\caption{
\textit{Spitzer}/IRAC image of M33 at 3.6~$\mu$m in units of MJy/sr.
 This image was created by co-adding the first four 
epochs of data. The emission traces the spiral morphology of the
galaxy and shows the stellar component of the galaxy reaching the
outer regions of the image ($R\simeq8.5$~kpc). The nucleus is
apparent as well as the largest star forming region in M33, NGC~604.
The outer ellipse
corresponds to the Holmberg radius with a semi-major axis of 8.7~kpc
(1\fdg2$\times$0\fdg6), and the inner ellipse bounds 90\% of 
the stellar population seen at 3.6~$\mu$m with a
semi-major axis of 7.5~kpc (1\fdg1$\times$0\fdg5). The regions used
for the study of Galactic foreground stars (\S\ref{field_contam}) are
highlighted with circles around the outer edges.  Areas delimited by 
blue contours were excluded from the variability search because the galaxy
surface brightness is too high. Point sources highlighted in green correspond 
to sources identified as YSOs.} 
\label{fig:3.6}
\end{figure}

\begin{figure}
\plotone{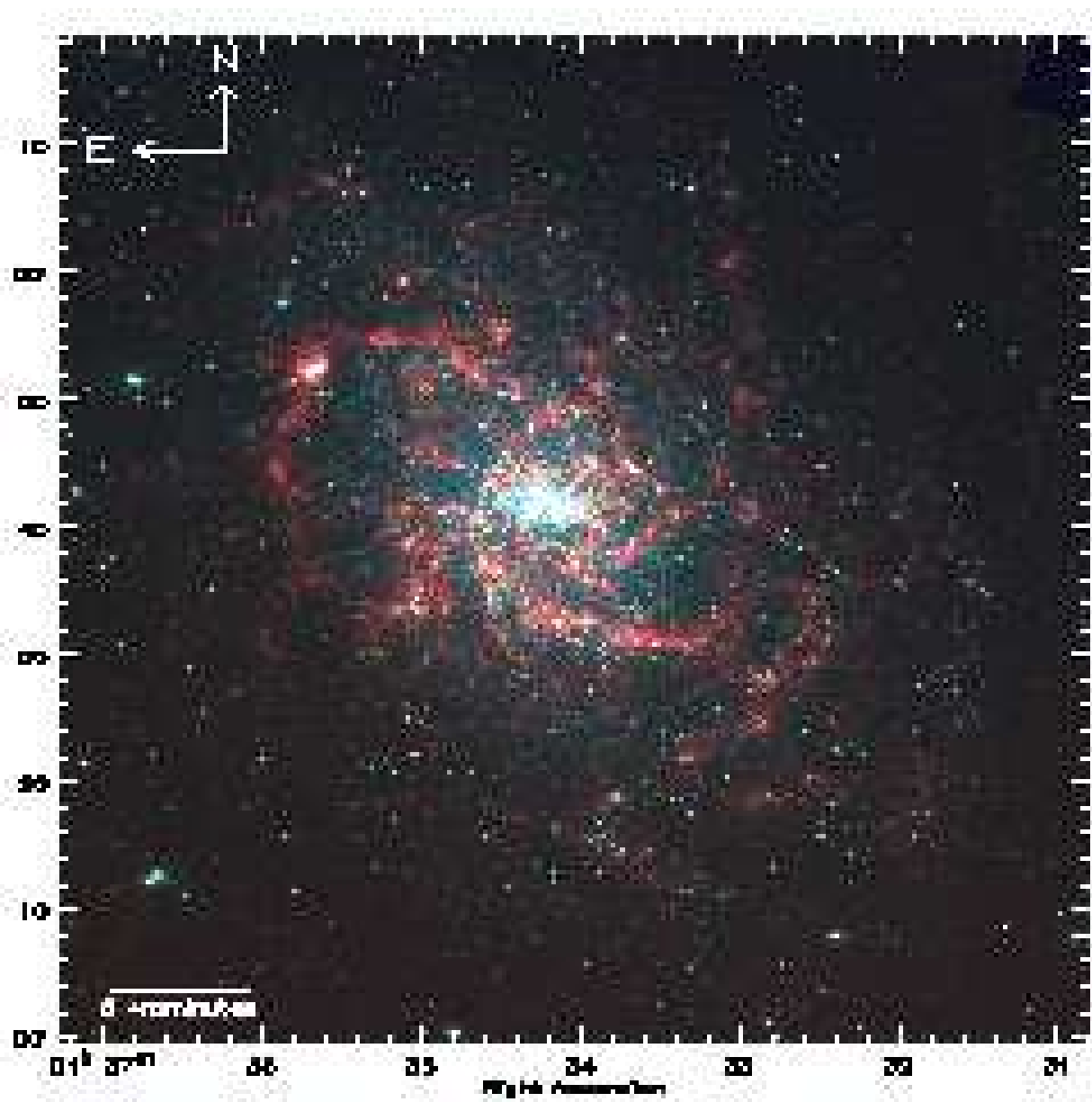}
\caption{Three-color IRAC image of M33 from co-adding all five epochs of 
data.  The 3.6, 4.5, and 8.0~$\mu$m mosaics are shown as blue, green, and red,
respectively.}
\label{fig:3color}
\end{figure}

This paper presents a five-epoch $\simeq$1.0~deg$^2$ IRAC survey of the
entire M33 galaxy.  Figures~\ref{fig:3.6} and~\ref{fig:3color} show
large scale IR maps of M33 at 3.6, 4.5, and 8.0~$\mu$m.  The
five epochs span a 19 month period (Table~\ref{tab:obs_summary}),
providing a unique opportunity to identify M33's variable star
population at IR wavelengths and determine their infrared colors and
spatial distribution.  Our survey covers an area similar to the CFHT
survey \citep{Har06}, and the resulting catalog of 2,923 variable
stars makes our work the first large scale variability study of M33 in
the IR.  Section~\ref{observ} describes the observations and the
data reduction process, \S\ref{field_contam} discusses contamination
by foreground Milky Way stars, \S\ref{variable} describes the
criteria used to identify variables, and \S\ref{properties} presents 
an overview of the classes of stars detected and their properties. 
The last section (\S\ref{conclusions}) gives a brief summary of 
our survey results.

\section{Observations \label{observ}}

M33 was mapped at six epochs with IRAC on \textit{Spitzer} as part of our 
Guaranteed Time Observer (GTO) program (Program ID~5, PI: R.~D.~Gehrz).  
The map at each epoch consisted of two consecutive astronomical observing 
requests (AORs) at all four IRAC wavelengths (3.6, 4.5, 5.8, and 8.0~$\mu$m).
The mapping sequence for each epoch consisted of $\simeq$148 positions 
per channel.
Each position was observed with three 12~s 
frames dithered with the standard, small, cycling pattern.
 The 4.5~$\mu$m and 8.0~$\mu$m images 
of any tile-location were obtained simultaneously. The 3.6~$\mu$m image 
was obtained a few minutes prior to or after the longer wavelength images. 
The FWHM of the point spread function at 3.6~$\mu$m is $1\farcs7$ or 
6.7~pc at the distance of M33. This specific 
study analyzes only the first five epochs, which cover a span of 19 
months, at 3.6, 4.5, and 8.0~$\mu$m. The sixth epoch was not available when 
the bulk of this present study was completed and is therefore 
not part of these results. The 5.8 $\mu$m images are generally 
similar to the 8.0~$\mu$m images but have lower sensitivity and were 
therefore excluded. 

The images were processed at the Spitzer Science Center (SSC) in the
Basic Calibrated Data (BCD) pipeline 11.0.2 (S11) which was the lastest 
pipeline version available at the time image processing was performed. 
Post-BCD processing was
carried out using an artifact mitigation algorithm developed by
\citet{Car05}. This algorithm alleviates the effects of muxbleed,
column pulldown/pullup, electronic banding, and bias variations
between the images. The results of the artifact mitigation processing 
showed no artifacts were introduced into the images. 
The SSC MOPEX software (2005 Sep 30 version,
\citealt{Mak05}) was used to mosaic the post-BCD images together. The
MOPEX processing includes background matching, outlier detection and
rejection, mosaicing, and coaddition. Background matching was
performed by minimizing the pixel differences in overlapping areas. 
The effect of flux clipping by 
MOPEX \citep{Sch06} was not accounted for during the mosaic processing. 
This effect, if present in the mosaicked images, introduces 
an additional error of 5\% and in the worst possible cases $\simeq 10$\%.  
The maximum (10\%) flux clipping error would therefore affect the identification of 
the 37 lowest amplitude variable stars in our catalog  
(amplitude $< 1.2$ mag) representing 1\% of the variable stars. The 
variability of these sources may be spurious and an artifact 
of processing. However, even if this effect is present, the 
vast majority of variable stars identified in 
our catalog show greater flux variability than 10\%, and their 
identification and characteristics are not affected. 
In the final step, each image was regridded onto a single world
coordinate system with a pixel scale of 1\farcs22 pixel$^{-1}$,
corresponding to 4.9 pc at the adopted distance of M33.
The final mosaics span an area of $\sim$1.0$^\circ$ $\times$
1.2$^\circ$ with the exact coverage at the outer edges varying
slightly from epoch to epoch.

Photometry of the stellar population was performed on mosaics using
the software package DAOphot/ALLSTAR, a point spread function (PSF)
based photometry package designed to handle high density star fields
and reject extended sources \citep{Ste87}.  The point source fluxes
were found by fitting a PSF generated in DAOphot using
bright, isolated stars in areas of low and flat background from
the mosaicked images of each channel. Due to DAOphot memory constraints, 
the M33 mosaics were each processed separately in two halves,
and more than 70 bright and isolated stars were selected in each
half to generate the PSFs. Other photometric procedures were tried
but produced higher uncertainties and proved less successful in 
recovering faint stars. Analysis of the residual images showed
that PSFs from a large mosaic are superior to ones generated from
individual frames. The individual images have lower signal to noise
than the 3-point dithered mosaicked image and have many fewer stars.
As a result, the PSF cannot be as well sampled as in the mosaic
image. Photometry conducted on smaller mosaics also yielded inferior
results, despite the smaller mosaics having more uniform stellar
densities and background. The smaller mosaicked images did not
contain a sufficient number of isolated stars in areas of low and
constant background from which the PSF could be generated.

The DAOphot photometric zero points were verified by aperture photometry of 
bright, isolated stars according to the prescription 
given by \citet{Rea05}. The 
DAOphot and aperture magnitudes agree within a dispersion of 0.07 mag at 
3.6~$\mu$m and 0.06 mag at 4.5 and 8.0~$\mu$m. These dispersions give 
estimates of the uncertainty of the DAOphot zero point. The true offset is 
likely to be smaller because some of the dispersion comes from errors in 
the aperture magnitude sky levels caused by confusion in the crowded 
fields. The slightly larger dispersion at 3.6~$\mu$m, where crowding is 
most severe, is probably a sign that confusion is important. Zero 
point offsets are likely to be similar at all wavelengths, leaving IRAC 
colors less affected than individual magnitudes, and of course zero point 
offsets cannot affect variability measures at all. The magnitudes given 
in this paper are thus on the IRAC S11 scale as described in the IRAC Data 
Handbook\footnote{http://ssc.spitzer.caltech.edu/irac/dh/iracdatahandbook3.9.pdf 
V3.0} within the stated uncertainties. 

\begin{figure}
\plotone{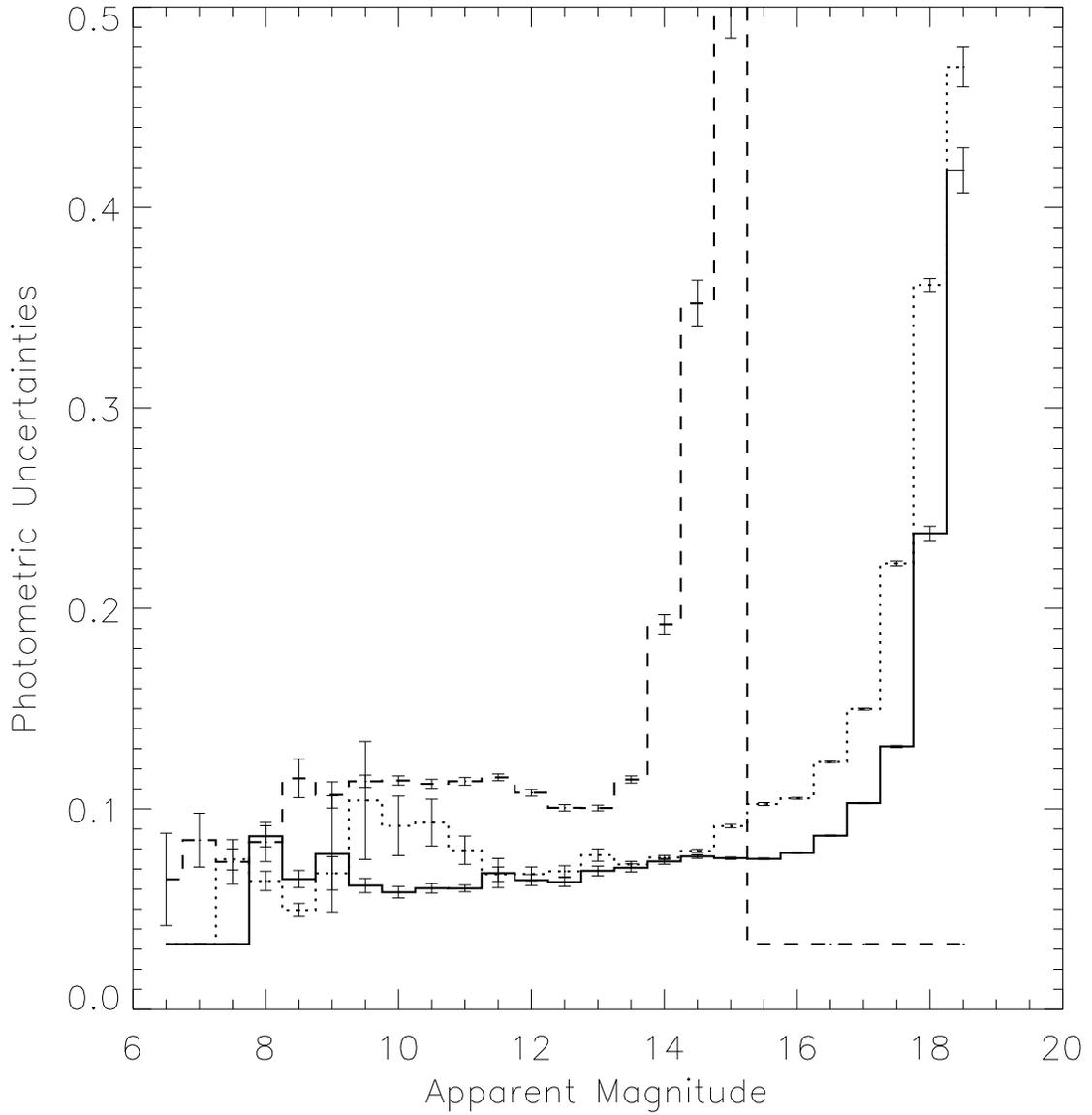}
\caption{Internal uncertainties ($\sigma_{int}$) for epoch~1. The  3.6,
4.5, and 8.0~$\mu$m mean uncertainties are shown as 
solid, dotted, and dashed lines,
respectively. Error bars show the uncertainty in the noise estimates
themselves. The noise properties of the other four epochs are 
consistent with these shown here.
\label{fig:noise}}
\end{figure}

DAOphot creates photometry lists with image pixel coordinates, fluxes, and 
accuracy parameters quantifying goodness of fit and sharpness (an image 
quality diagnostic). DAOphot provides an uncertainty estimate that 
includes detector readout noise, photon statistics, accuracy of the 
PSF-fit, and uncertainty in the PSF itself. The final internal uncertainty 
($\sigma_{int}$) for each epoch includes an additional 0.03~mag added in 
quadrature to account for the dispersion in repeated measurements of any 
star \citep{Rea05}. For stars in the final catalogs, the median photometric 
uncertainty (Figure~\ref{fig:noise}) is 10\%, the minimum is 5\%, 
and 90\% of the measurements have uncertainties $<$20\%. 

\begin{figure}
\plotone{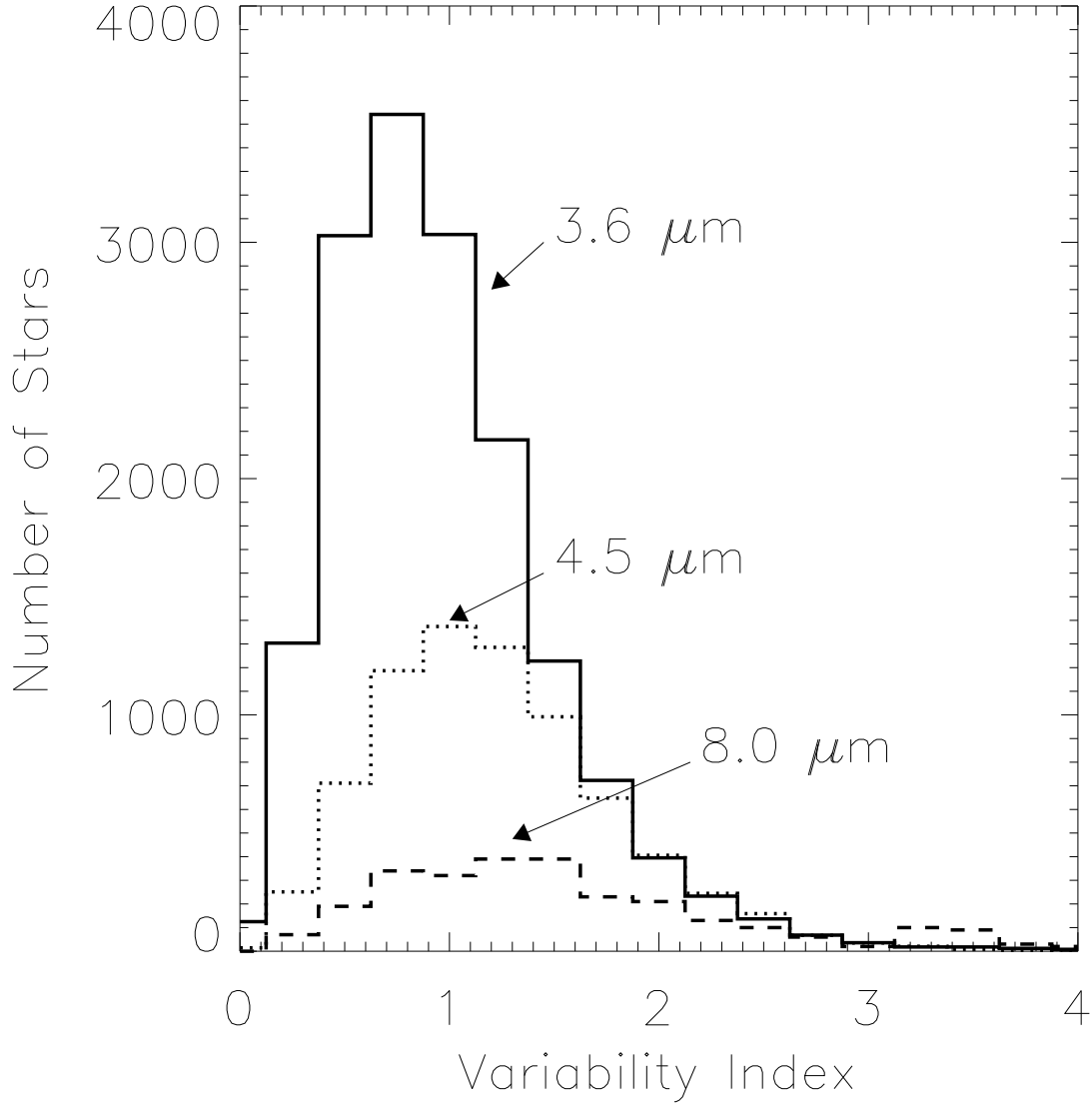}
\caption{Variability index histogram for stars in the faintest one
magnitude above the completeness limit 
at each wavelength (\S\ref{variable}):
3.6~\micron\ (solid line), 4.5~\micron\ (dotted line), and 8.0~$\mu$m
(dashed line).  The 8.0~\micron\ histogram has been multiplied by ten. 
At each wavelength,
the histogram includes only stars with measurements at all five
epochs. Variability index is defined \citep{Gal04} as the
ratio of standard deviation of the measurements at the five epochs to
the mean internal uncertainty $\sigma_{int}$. 
Because the internal uncertainties are nearly constant for a
given star, the variability index distribution will be close to a
chi-square distribution except for the variables.  The tail at large
variability indices indicates the variable stars.}
\label{fig:chi2_distrib}
\end{figure}

The individual uncertainty estimates were verified by analyzing the 
observed discrepancy in flux for each star from epoch to epoch.
Variability index  \citep{Gal04} is defined as the
ratio of standard deviation of the measurements at the five epochs 
for a star to the mean internal 
uncertainty $\langle\sigma_{int}\rangle$ of that star. 
Because the internal uncertainties are nearly constant for a
given star, the variability index distribution (Figure~\ref{fig:chi2_distrib}) 
will be close to a chi-square distribution except for the presence of 
variable stars in the tail of the distribution.
Nearly 95\% of sources at 3.6~$\mu$m, 91\% of sources at 
4.5~$\mu$m, and 78\% of sources at 8.0~$\mu$m cluster 
below a variability index 
value of three at each wavelength band 
suggesting that although the flux measurements do vary between epochs, the 
dispersion is less than 3$\sigma$ of the $<\sigma_{int}>$. 

A master star catalog was created from the 15 individual photometry lists 
(five epochs, three wavelengths each) using DAOMATCH and 
DAOMASTER. DAOMATCH and DAOMASTER  are separate programs 
that merge photometry output files 
from DAOphot into a single file with matched point sources. DAOMATCH finds 
the translation, rotation, and scaling solution between the photometry 
files, and DAOMASTER then matches point sources by pixel coordinates. The 
final coordinates were converted to right ascension (RA.) and declination 
(Decl.) using the WCSTools' xy2sky coordinate conversion 
program\footnote{http://tdc-www.harvard.edu/software/wcstools/}. A total 
of over 100,000 sources were cataloged. When the fluxes from all of these 
stars are summed, the total flux corresponds to a 3.6~$\mu$m
apparent magnitude of $[3.6] = 3.7$.  This value is comparable to 
the total M33 3.6~$\mu$m apparent magnitude of $[3.6] = 3.2$ obtained 
from surface brightness fitting. We conclude that $\simeq$ 60\% of the 
3.6~\micron\ flux of M33 can be resolved into point sources 
at {\it Spitzer} resolution. 

\begin{figure}
\plotone{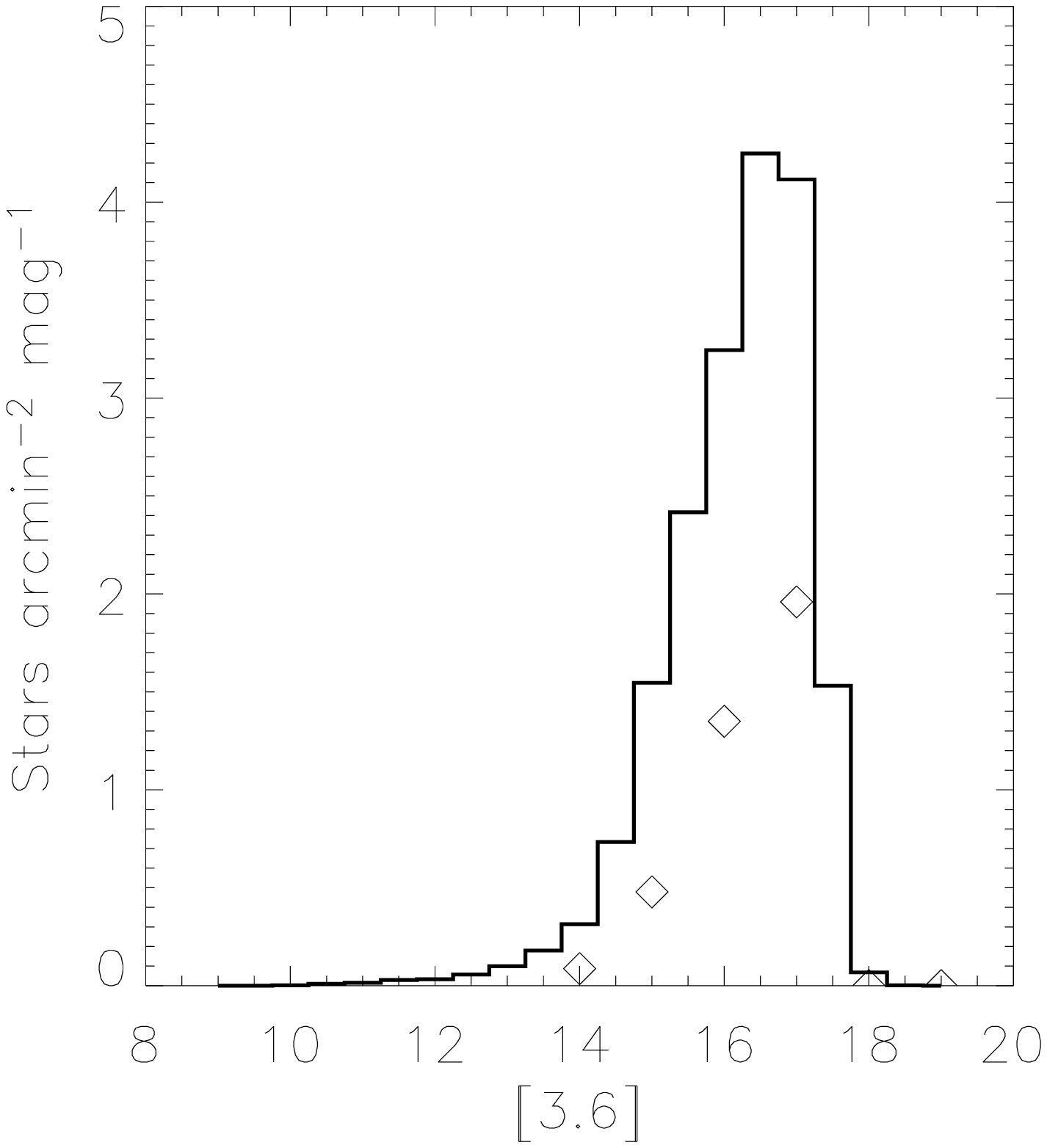}
\caption{The 3.6~$\mu$m normalized star counts. The solid 
histogram represents counts in the entire
catalog (excluding high surface brightness regions),  and 
diamonds represent counts from the outer fields with bin sizes twice 
the width of the catalog's bin sizes.}
\label{fig:field_lum3.6}
\end{figure}

Detailed completeness limit analysis was difficult given the wide
range of stellar density conditions found across our field of
view. However, the completeness limits can be estimated from the
cumulative luminosity functions (Figure~\ref{fig:field_lum3.6}) and
the measured noise levels (Figure~\ref{fig:noise}). We estimate that our 
catalog is complete to 16.6 mag at 3.6~$\mu$m, 16.6 mag at 4.5~$\mu$m, and 
14.6 mag at 8.0~$\mu$m.  The expected IRAC 5$\sigma$
sensitivity at 3.6~\micron, calculated with the {\sc senspet}
tool\footnote{http://ssc.spitzer.caltech.edu/tools/senspet} and assuming a 
high background level (0.25~MJy/sr) is 18.3 mag.  The difference between the 
expected sensitivty of 18.3 mag and the estimated completeness limit 
of 16.6 mag probably arises from the blending of sources in the 
crowded regions of M33.  In fact, in the
most crowded central regions, reliable photometry is quite
difficult. Although cataloged, stars in regions of 3.6~\micron\ surface
brightness $>$1.06~MJy/sr were not examined for variability.
(Isophotes of constant surface brightness were found after smoothing
the 3.6~$\mu$m epoch~1 image with a 20-pixel-radius Gaussian kernel.)
The surface brightness cutoff corresponds to the central region of
the galaxy ($R_{gal} < 0.4$~kpc) and the center of bright
star-forming regions such as NGC 604.  A total of 934 stars were
found in these regions of high surface brightness.

\begin{figure}
\plotone{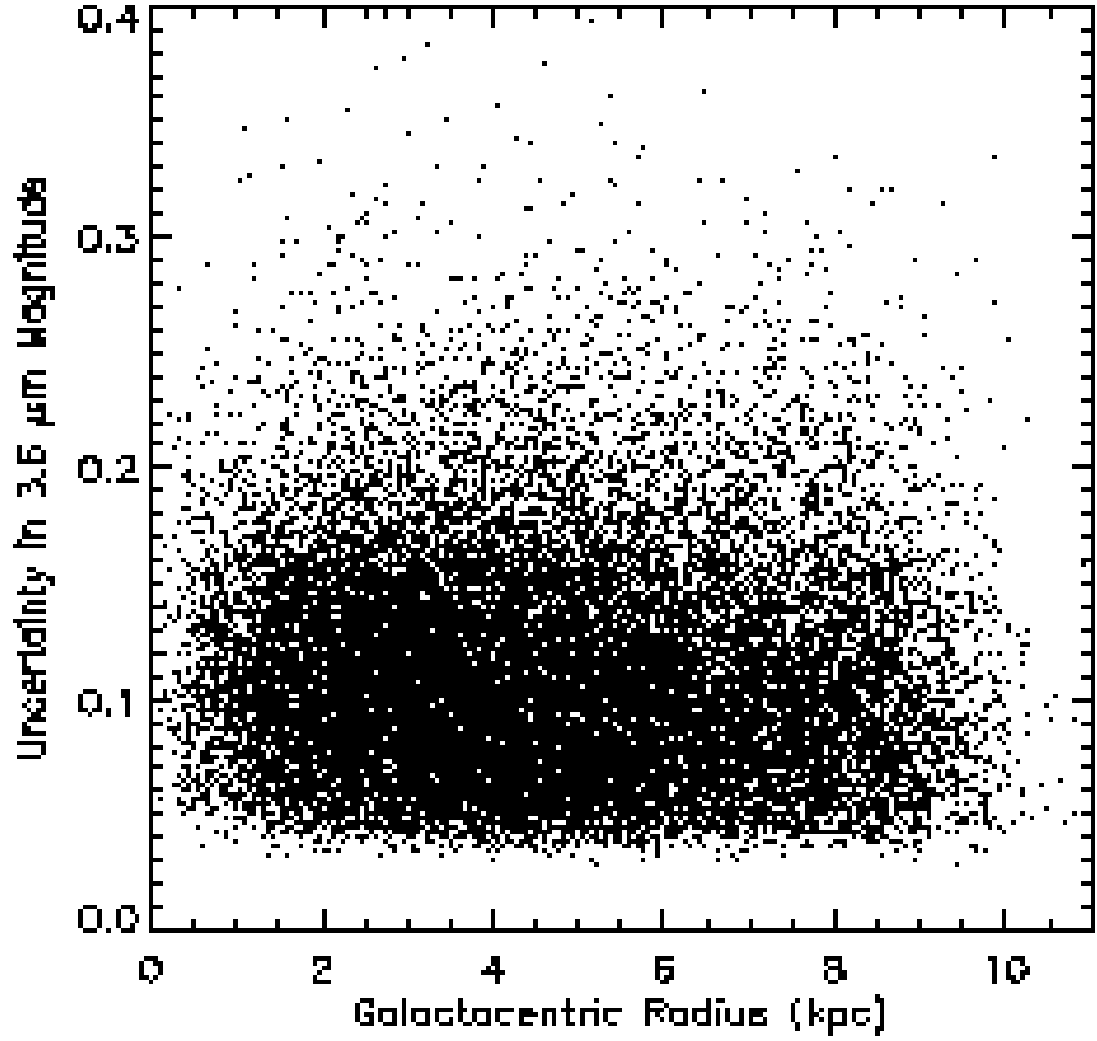}
\caption{The distribution of the uncertainty in the 3.6~$\mu$m magnitude
is constant across galacto-centric radii.}
\label{fig:errors_rad}
\end{figure}

The final catalogs consist of stars detected at the first 
five epochs in at least one wavelength. The 
catalogs are given in Tables~\ref{tab:catalog1} (non-variable)
and Tables~\ref{tab:catalog2} (detected variables, see \S\ref{variable}).
Complete tables are available in electronic form in the online 
version of this paper accessible through the Astrophysical Journal. 
These tables contain 37650 
sources, which can be referred to by the acronym `SSTM3307' followed by 
the source's coordinates as indicated in the catalog\footnote{eg., SSTM3307 
J013151.74+302545.2      22.965624      30.429226}.  Because of the 
lower sensitivity of IRAC at 8.0~$\mu$m and the Rayleigh-Jeans spectrum of 
most stars, only 5,537 stars were detected even once at 8.0~$\mu$m, and 
only 2,689 were detected at all five epochs.  Stars in the high surface 
brightness regions are included in the non-variable catalog 
for convenience, but they 
were not tested for variability because the indicated magnitude 
uncertainties ($\sigma_{int}$) may be underestimated.  With this exclusion 
of the regions of high-surface brightness, 
the uncertainties do not change much with galacto-centric radius, as shown 
in Figure~\ref{fig:errors_rad}. 

\section{Discussion \label{discuss}}

\subsection{Galactic Foreground Star Contamination \label{field_contam} }

The M33 IRAC observations extend out to the
8.7 kpc Holmberg radius \citep{Hol58} along the semi-major axis of
the galaxy but not beyond.  This limits the area available for a
field contamination analysis.  Moreover, each IRAC epoch has slightly
different areal coverage at the edges, further reducing the area
covered at all five epochs. Using the limited areal coverage 
available in the images, two methods were used to estimate the 
field contamination, namely a first order determination of the IR stellar 
component extent and an analysis of selected 
fields near the image limits.

\begin{figure}
\plotone{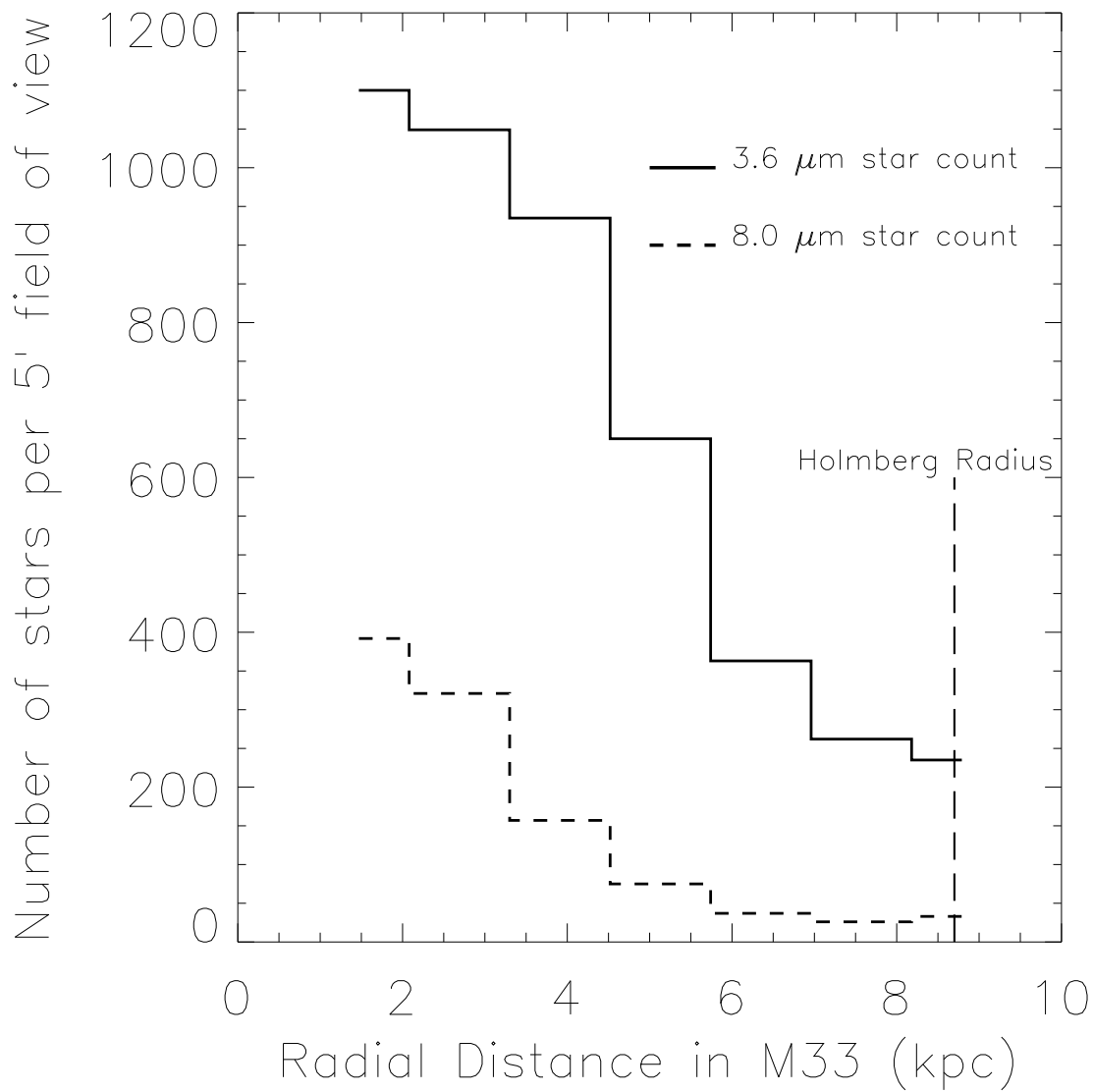}
\caption{The radial distributions of the stellar component found in
$5^\prime\times5^\prime$ fields along the semi-major axis of M33 at
3.6 and 8.0~$\mu$m in epoch 1 are plotted to the areal limits of the
IRAC observations reaching the Holmberg radius (R = 8.7 kpc) of the
galaxy. The stellar counts flatten at 7.5 (6) kpc at 3.6
(8.0)~$\mu$m. The 4.5~$\mu$m star counts are similiar to the 3.6~$\mu$m 
counts.}
\label{fig:holmberg_dist}
\end{figure}

Our first method examines to first order the extent of the IR
 stellar component in M33 by analyzing the
 radial distribution of point sources. Radial star 
counts across the semi-major axis (Figure~\ref{fig:holmberg_dist}) show 
that 90\% of the stellar population is contained within a galacto-centric 
radius of 7.5 kpc at 3.6~$\mu$m, and 95\% of stars are within 6.0 kpc at 
8.0~$\mu$m. The flattening of the star counts past $R\simeq 7.5$~kpc 
indicate that field stars from the Milky Way become important at that 
radius.

\begin{figure}
\plotone{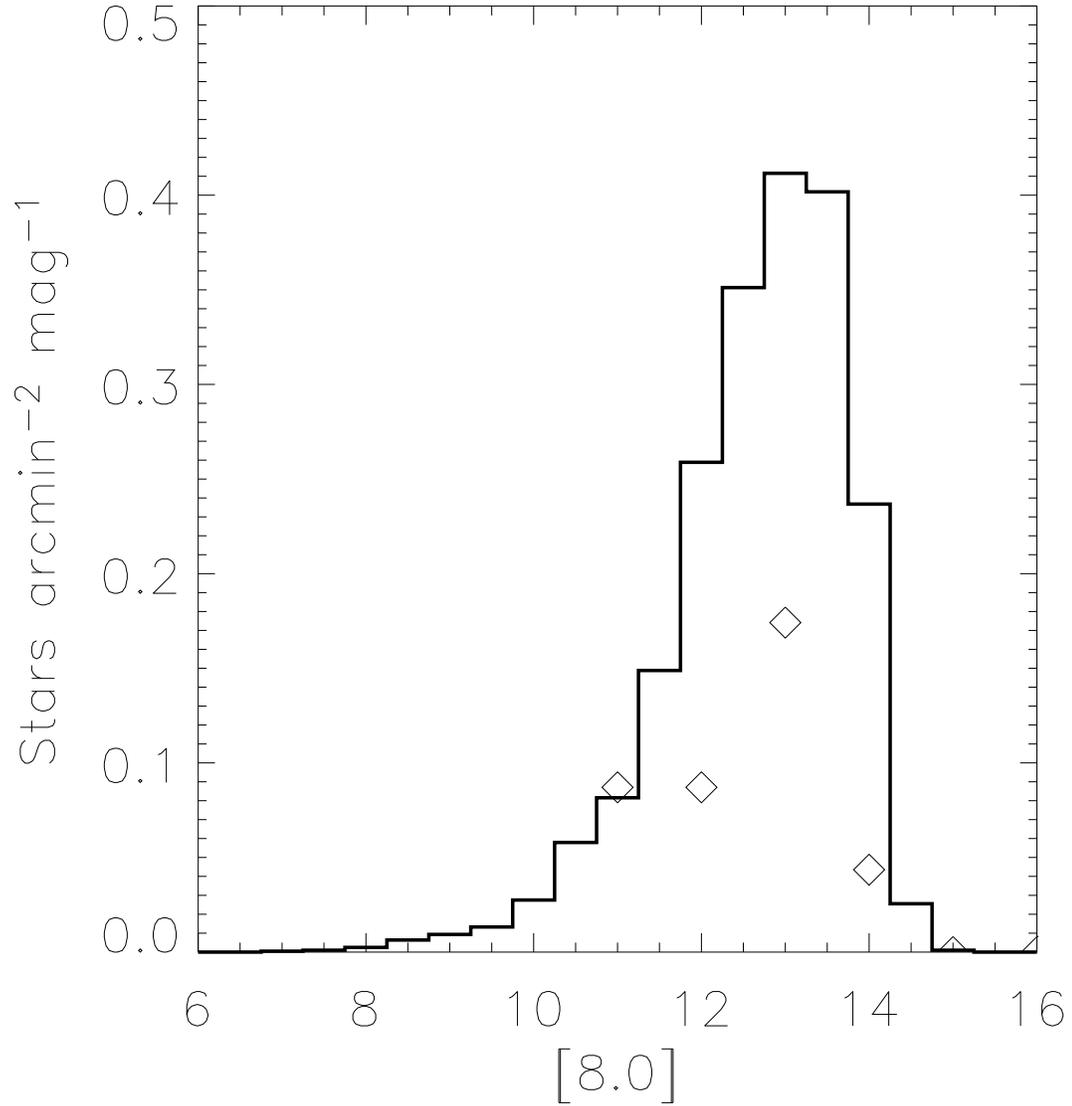}
\caption{The 8.0~$\mu$m normalized star counts. The
solid histogram represents counts in the entire catalog (excluding
high surface brightness regions), and diamonds represent counts from
the outer fields used for the contamination study with bin sizes 
twice the width of the catalog's bin sizes.}
\label{fig:field_lum8.0}
\end{figure}

The radial distribution of stars was used to guide the selection of 
thirteen reference fields located outside the Holmberg radius used 
to estimate the field contamination in our second method.
Each field had a 
radius of $0\farcm75$ for a total of 23~arcmin$^2$; the regions are shown 
in Figure~\ref{fig:3.6}.  At 3.6~\micron, 78 catalog stars lie 
within these regions for an indicated stellar 
density of 12,200 stars~deg$^{-2}$.  Only 7 stars were 
identified at 8.0~$\mu$m making accurate extrapolation difficult. The 
star count at 3.6~$\mu$m in the outer fields corresponds to 
6,900 stars or about 20\% of the total within an ellipse 
at the Holmberg radius (0.57~deg$^2$). This 
should be an upper limit on catalog stars that do not belong to M33. 
For comparison, the DIRBE Faint Source Model \citep[][FSM]{Are98} predicts 
about 6,400 stars~deg$^{-2}$ within the 3.6~$\mu$m magnitude range of our 
survey.  The difference in the observed stellar areal density 
(12,200~deg$^{-2}$) and the predicted field star areal density 
(6,400~deg$^{-2}$) suggests that even in the 
outer regions of the image about half the catalog stars belong to M33, 
and the contamination within the Holmberg radius is only about 10\%.  
Because the stars have a relatively small range in colors, the 
contamination percentage is about the same at all three catalog 
wavelengths. Figures~\ref{fig:field_lum3.6} and \ref{fig:field_lum8.0} 
compare the luminosity functions of stars found throughout M33 to stars in 
the selected outer regions. 

\begin{figure}
\plotone{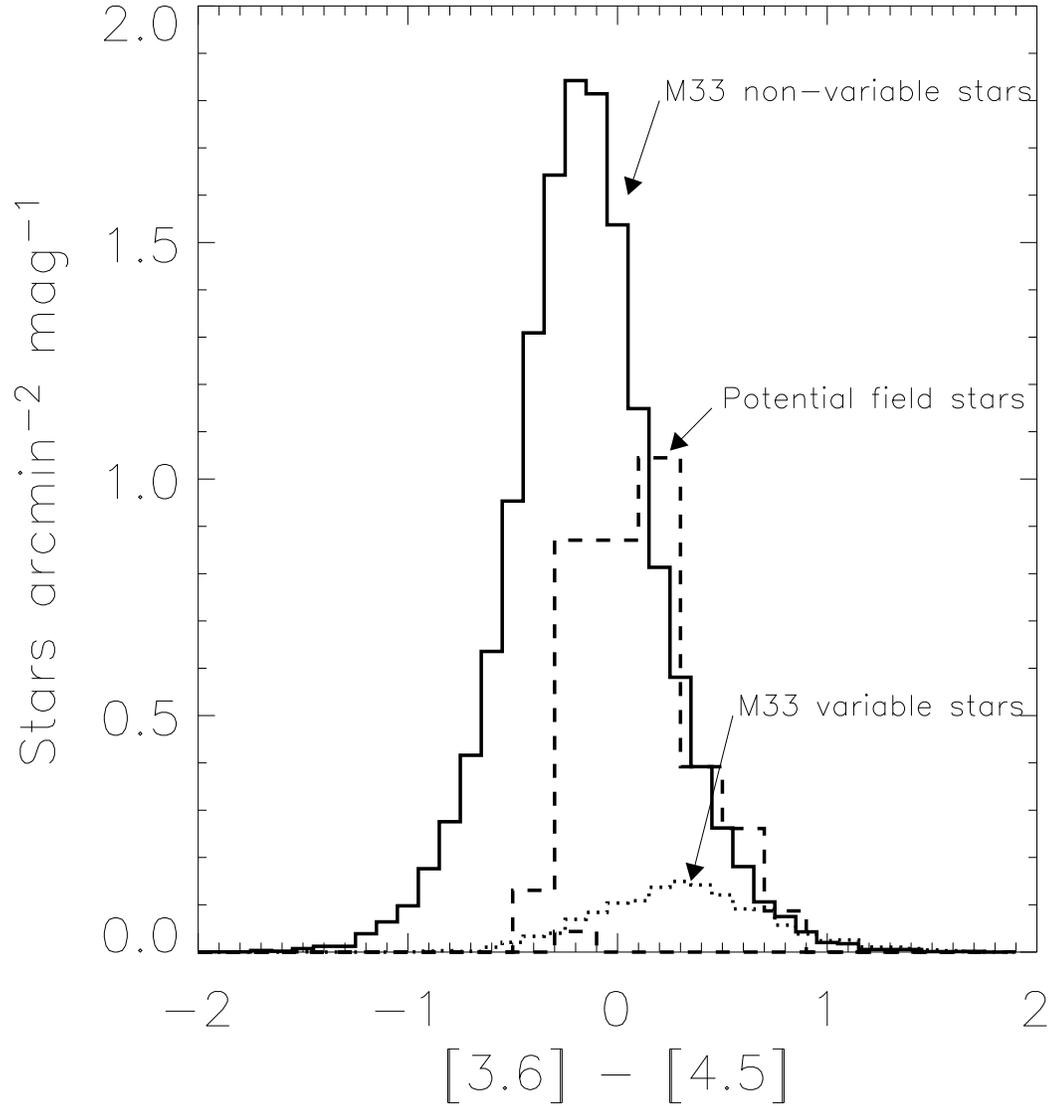}
\caption{Normalized star counts versus color. The solid line shows
the color distribution for all non-variable stars in the catalog
(excluding high surface brightness regions), the dotted line
shows the variable stars, and the dashed line shows the colors
of stars in the outer fields binned with 2$\times$ 
catalog's bin size.}
\label{fig:field_colors12}
\end{figure}

\begin{figure}
\plotone{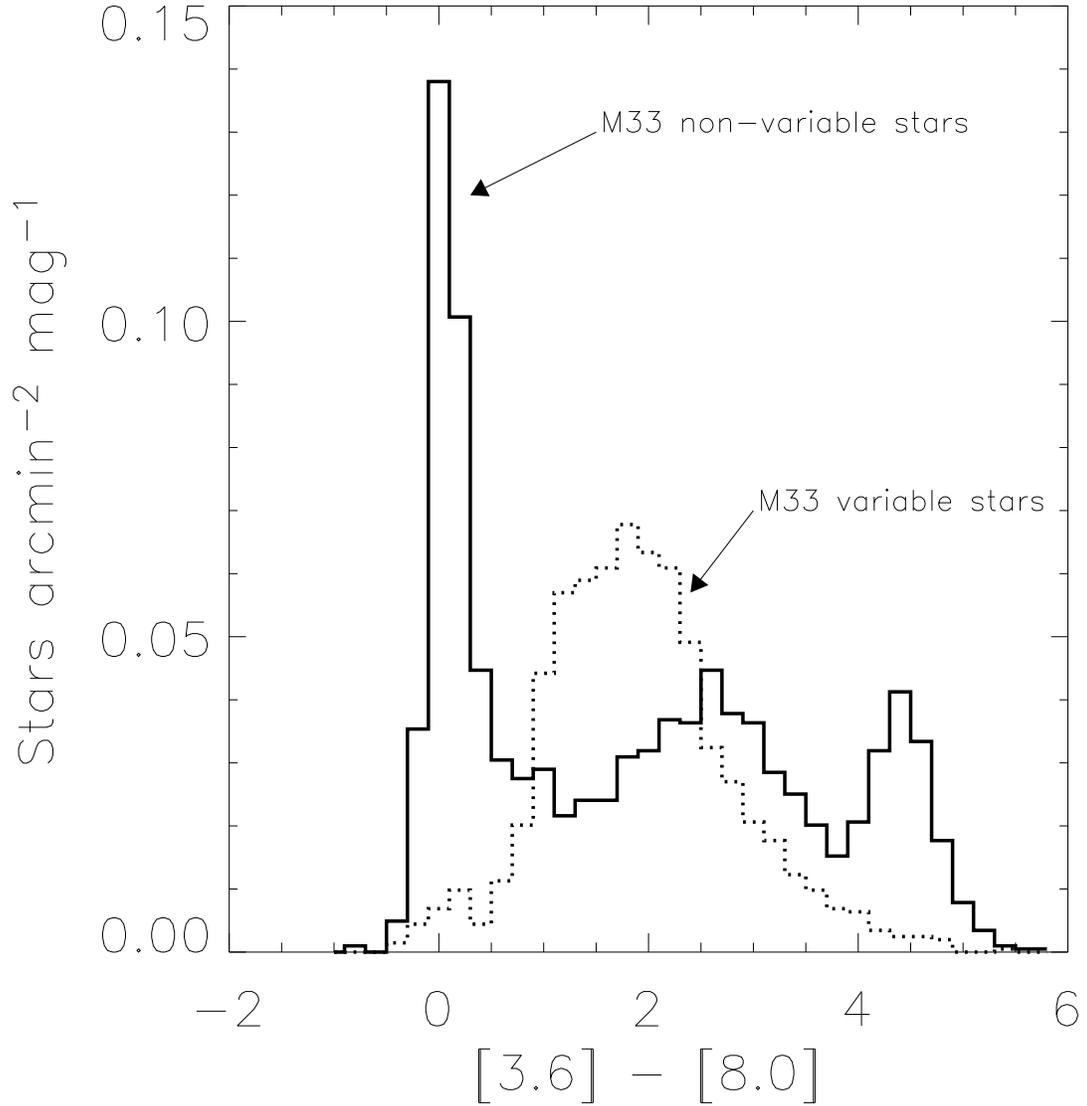}
\caption{Normalized star counts versus color.  The solid line shows
the color distribution for all non-variable stars in the catalog
(excluding high surface brightness regions) and the dotted line
shows the variable stars. The field regions had a median $[3.6] - [8.0]$ 
color of 2.3 with a range from 0.1--3.2 but had insufficient number 
of stars to bin.}
\label{fig:field_colors14}
\end{figure}

Foreground stars must be main sequence (``dwarf'') stars because a
giant star in the Milky Way would have $[3.6] \ltsimeq 7$ mag, much
brighter than any of the sources in the catalog.  Dwarfs are redder
in $[3.6] - [4.5]$ color than giants because they have water vapor
absorption at 3.6~$\mu$m and lack CO absorption at 4.5~$\mu$m
\citep{Mer76}.  Figures~\ref{fig:field_colors12} and
\ref{fig:field_colors14} show the color distributions of the catalog
stars.  As expected for foreground dwarfs, the stars in the outer
fields occupy the redder half of the $[3.6]-[4.5]$ color
distribution.  Regardless of the similarity in colors, the
contamination of the variable star catalog should be very small because
the foreground main sequence dwarfs are not variable. Indeed, in the
outer fields, only one star met our criteria for variability, and it
is likely an M33 member.  Therefore, although about 10--20\% of the
non-variable stars could be foreground point sources, the
contamination component to the identified variable population is
negligible.

\subsection{Variability Search \label{variable}}

Strict criteria for variability were chosen to avoid having spurious 
detector events interpreted as variablity in a star.  The 
disadvantage of this 
approach is that some true variables will not be flagged as such.  As 
noted in \S\ref{observ}, only stars detected at all five 
epochs in at least one wavelength are in the catalog.  
However, once a star makes it to the catalog all available measurements of 
that star are included.  The variability search was limited to stars with 
mean magnitudes brighter than the estimated completeness limits (16.6, 
16.6, and 14.6 mag at 3.6, 4.5, and 8.0~\micron, respectively) and stars 
outside regions of highest surface brightness (see \S\ref{observ}). 

\begin{figure}
\plotone{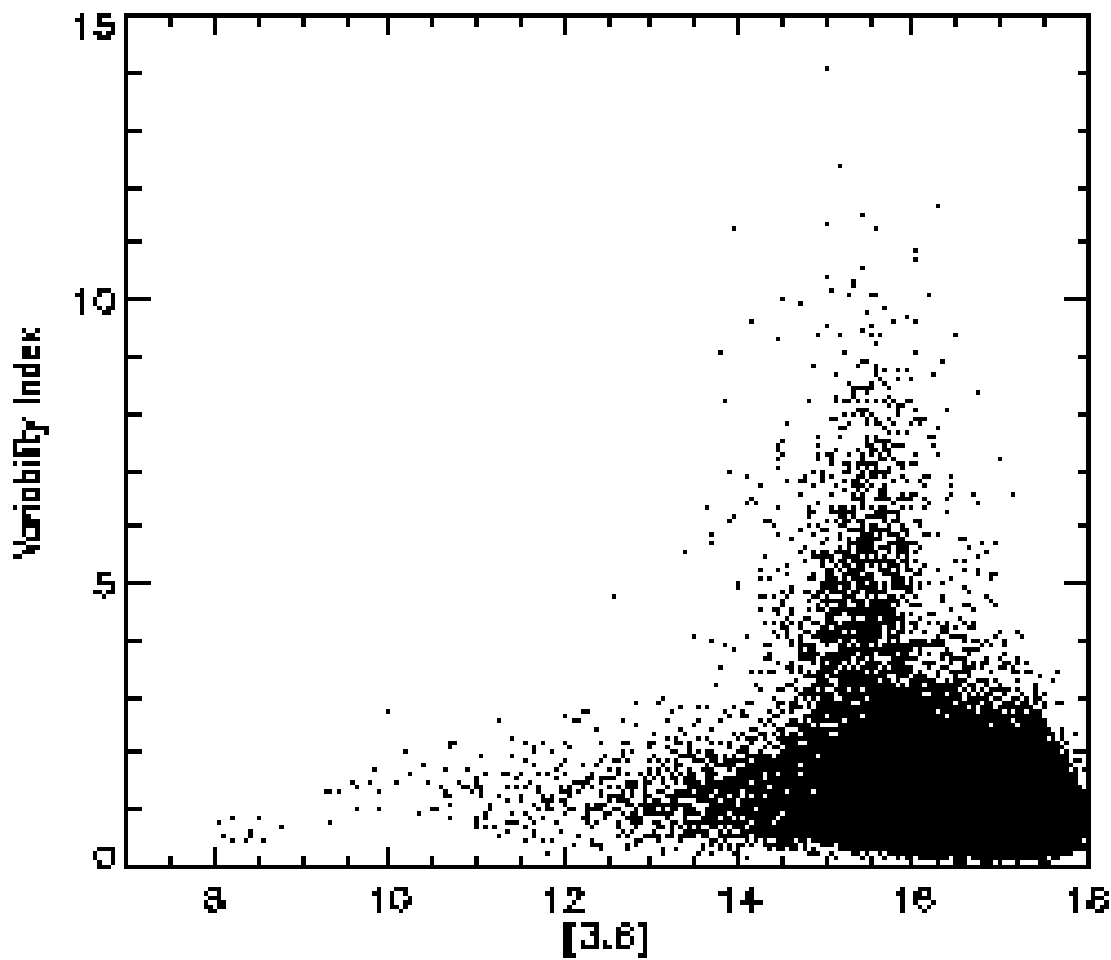}
\caption{Variability index at 3.6~$\mu$m versus magnitude (\S\ref{variable}). 
A total of 37,699 stars with 3.6~\micron\ detections at all five epochs are 
included. The
majority of sources lie below an index value of 3. Stars above this
threshold are considered variable provided the high variability is
not limited to a single epoch at a single wavelength.}
\label{fig:index_ch1}
\end{figure}

\begin{figure}
\plotone{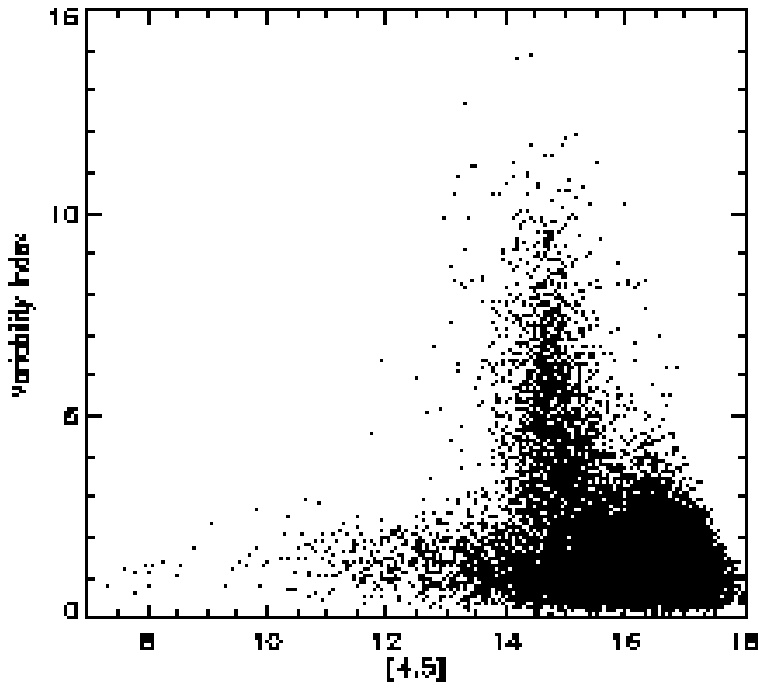}
\caption{Variability index at 4.5~$\mu$m versus magnitude (\S\ref{variable}). 
A total of 37,773 stars with 4.5~\micron\ detections at all five epochs are
included.  The same criteria for variability apply as at
3.6~\micron.}
\label{fig:index_ch2}
\end{figure}

\begin{figure}
\plotone{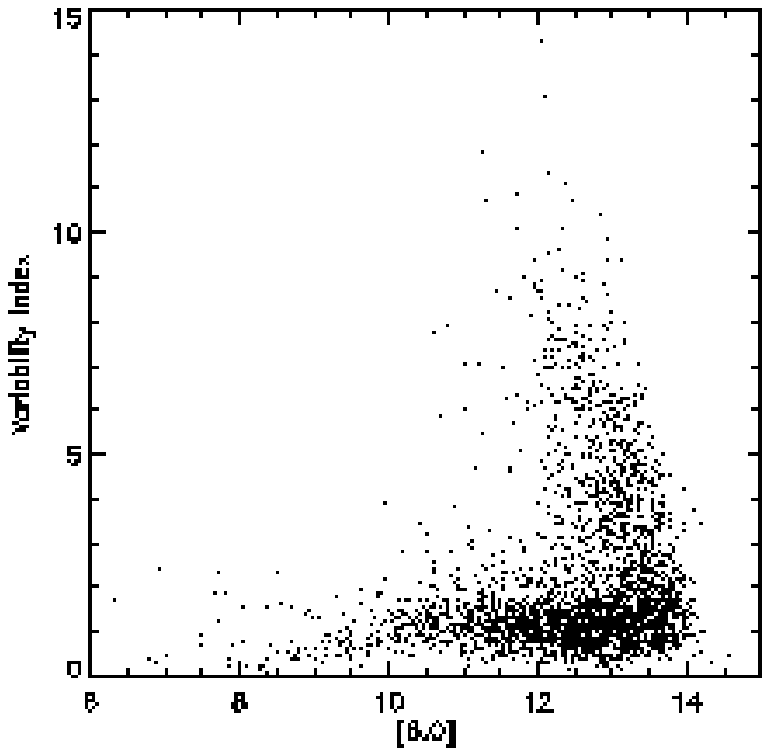}
\caption{Variability index at 8.0~$\mu$m versus magnitude (\S\ref{variable}). 
A total of 2,689 stars with 8.0~\micron\ detections at all five epochs are included.
The same criteria for variability apply as at 3.6~\micron.}
\label{fig:index_ch4}
\end{figure}

The threshold criterion for indentifying what 
magnitude variation could be attributed to intrinsic variability versus 
measurement fluctuations was determined empirically by comparing the 
dispersion in the magnitude at all epochs with the internal uncertainties 
of a given star (Figures~\ref{fig:index_ch1} --\ref{fig:index_ch4}, see 
variability index discussion in \S\ref{observ}). 
A minimum variability index value of three ensures conservative variable 
star identification and represents a deviation 
from a mean magnitude of at least 
0.1~mag at 3.6~$\mu$m depending on the point source's uncertainties. 
The variability search itself consisted of comparing the magnitude at an epoch 
when a star was detected to the mean magnitude at that wavelength. 
If the magnitude difference was at least $3 \times \sigma_{int}$ for that 
epoch, the measurement was flagged as indicating intrinsic variability.

The final catalog of variables consists of all stars 
flagged as variable in two or more 
measurements. (This discrepancy can include two wavelengths at a single 
epoch or two epochs at a single wavelength.)  
Requiring two discrepant measurements was necessary because 
a single discrepant magnitude can be caused by 
an unremoved cosmic ray or other 
processing artifact. Table~\ref{tab:num_var} shows 
the number of variable stars identified at each wavelength. 

\subsection{The Properties of Star Classes in the IRAC Bands}\label{properties}

\begin{figure}
\plotone{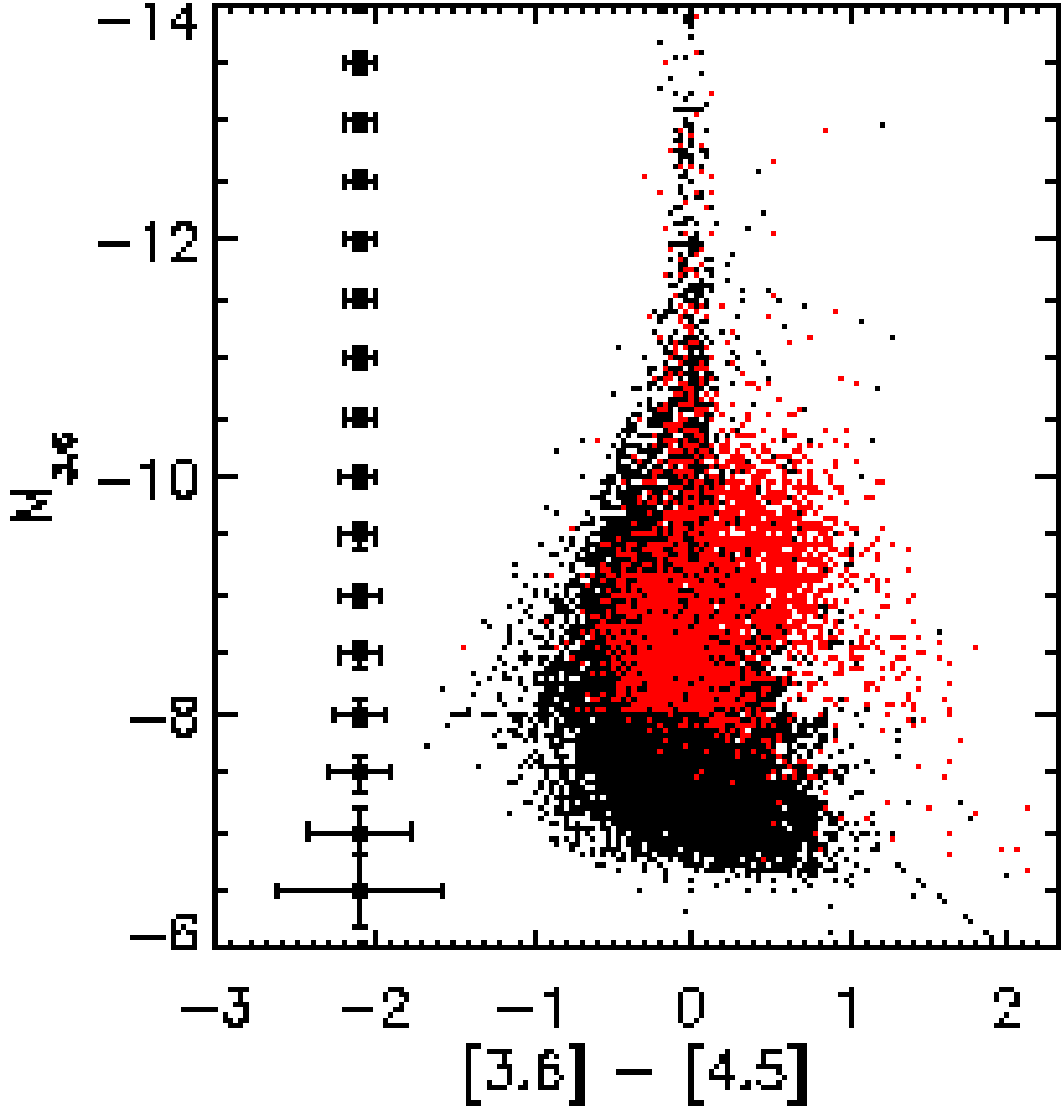}
\caption{Color-magnitude diagram showing 35,140 point sources 
identified in the catalog with measurements in 
all five epochs at 3.6 or
4.5~$\mu$m. Non-variable sources are plotted in black, variables in
red. The diagonal line shows the completeness limit of the
survey. The error bars show representative uncertainties of a single
epoch measurement; they do not vary significantly with color.
 The absolute magnitudes were calculated
assuming a distance modulus of 24.60 corresponding to the adopted
distance of 830~kpc.}
\label{fig:cmd12}
\end{figure}

\begin{figure}
\plotone{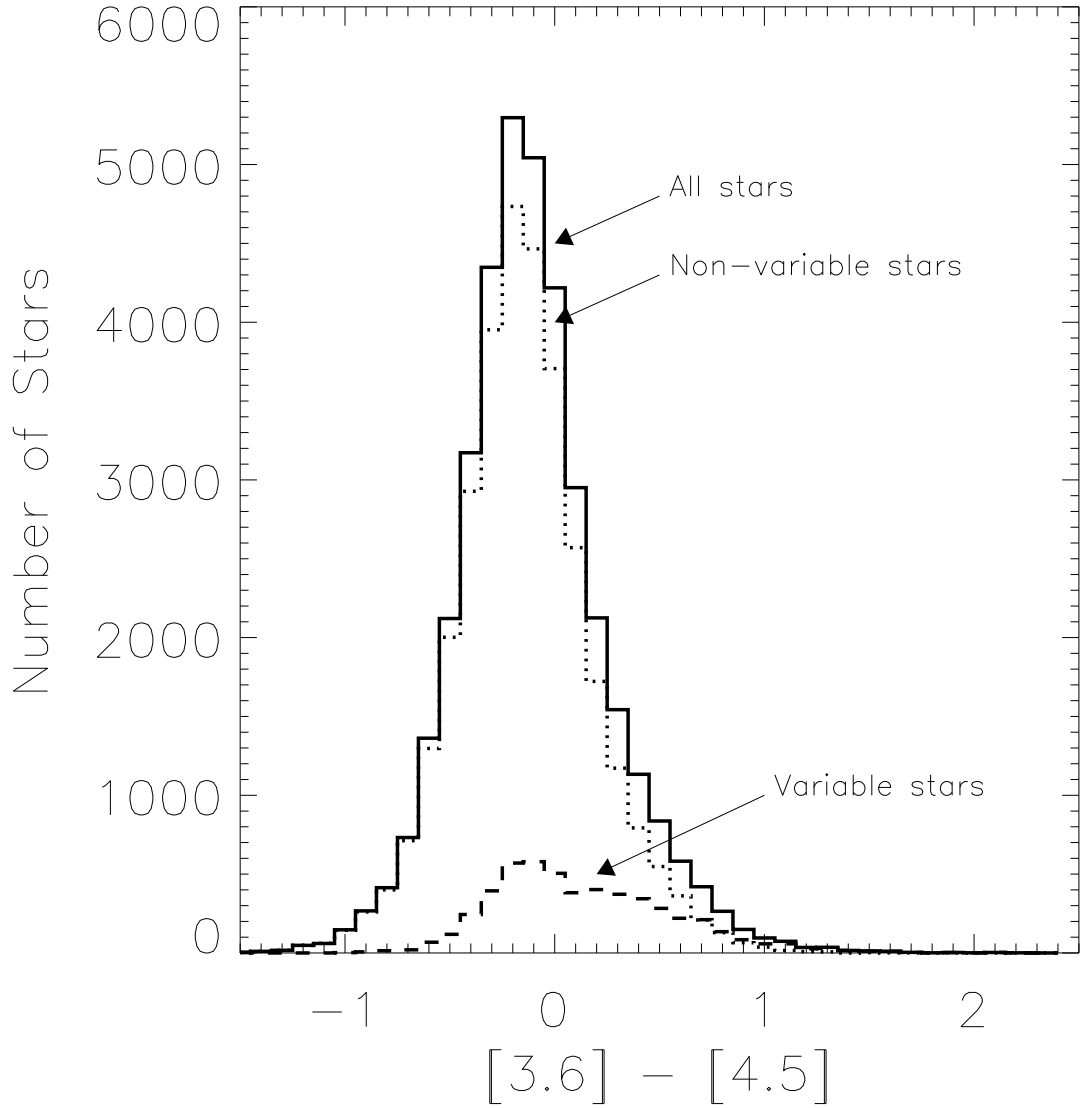}
\caption{The $[3.6] - [4.5]$ color distribution of non-variable
sources (dotted line), variable sources (dashed line), and both types
combined (solid line). The dip at $[3.6] - [4.5] = 0.2$ in the
distribution of variable stars indicates the separation between
oxygen-rich and carbon stars.}
\label{fig:color_histogm}
\end{figure}

Figure~\ref{fig:cmd12} shows a color-magnitude diagram (CMD) of all point 
sources in both catalogs.  The non-variable sources have colors centered on 
$-0.10$, as also shown in Figure~\ref{fig:color_histogm}. This slightly 
blue $[3.6] - [4.5]$ color is typical of oxygen-rich giant stars whose 
photospheric CO and SiO absorption bands fall within 
the 4.5~$\mu$m bandpass \citep{Mar07, Bol07}. The 
variable stars have a bimodal distribution. One peak is at $[3.6]-[4.5] 
\approx -0.10$, the same color as the non-variable, oxygen-rich stars. The 
second peak is at $[3.6] - [4.5] \approx 0.3$.  Stars in this color region 
are mostly carbon stars. The slightly red $[3.6] -[4.5]$ color is due to 
the strong photospheric absorption features of C$_2$H$_2$ and HCN 
in the 3.6~$\mu$m bandpass \citep{Mar07}. In this part of the CMD, 
34\% of the stars are flagged as variable.
Quite likely a much larger fraction of the 
carbon stars are variable but not detected as such with our strict 
criteria. 

In other Local Group galaxies, the tip of the red giant branch (TRGB) is
found at $M_{L'} \simeq -6.4$ for the Magellanic Clouds \citep{van05} 
and $M_{3.6} \simeq -6.6$ for the dwarf galaxy WLM
\citep{Jac07}.  While age and metallicity will have an impact of 
$\la$0.3~mag depending on the wavelength studied \citep{Cio03}, the
TRGB for M33 should be at $M_{3.6} \ga -7.0$ mag, below the completeness
limit of our survey. Although the observations lack 
sufficient sensitivity to detect TRGB 
stars, they do, however, have ample sensitivity to detect AGB stars. 
The observational limit at $M_{3.6} = -7.6$ mag at the distance to M33 
corresponds to the brightness expected of luminous AGB stars and red
 supergiants \citep{Cox00}. Dust modeling predictions for AGB
stars \citep{Gro06} give expected absolute magnitudes ranging from
$M_{3.6} \simeq -7.15$ mag for late M stars without appreciable mass loss
to M$_{3.6} \simeq -8.26$ mag for early C stars with mass-loss.
\citet{Whi2006} found $-8.5\ga M_{3.6}\ga-10$ for a sample of carbon
stars that should represent the most luminous ones in the Milky Way. 
A range of $2.5 \la M_{V}\la 10$ is typical for Miras and
$M_{V} \approx -1$ for semi-regulars and slow irregulars. 
The stars detected in our catalogs lie above the TRGB and are thus 
likely to be AGB stars.

\begin{figure}
\plotone{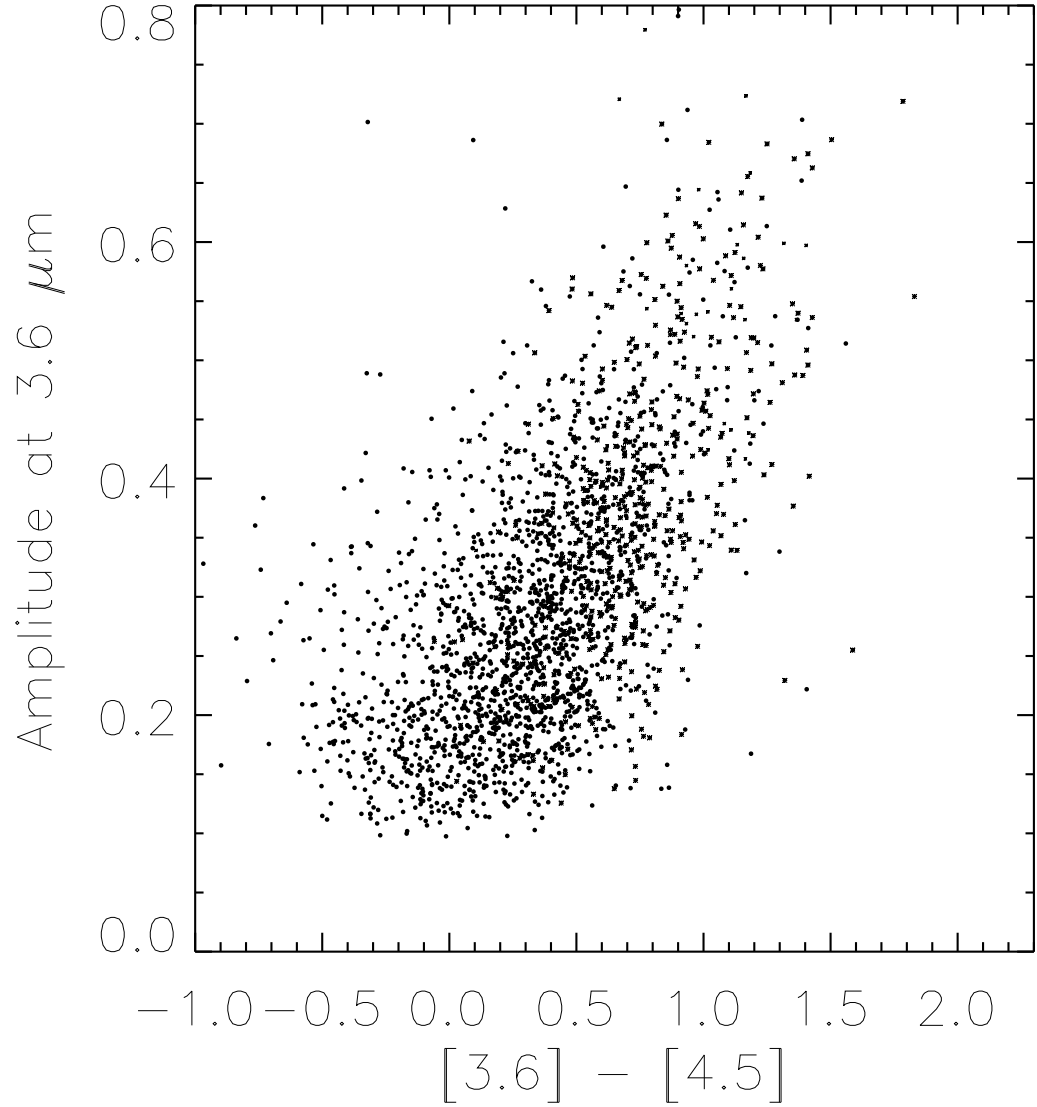}
\caption{Amplitude of variation versus $[3.6] - [4.5]$ color. The amplitude 
is defined as the standard deviation of the magnitudes for the five epochs 
of observation at 3.6~\micron\ for each star. The amplitude of these stars 
at 4.5 $\mu$m is similar with a range from 0.1 to 1.0. } 
\label{fig:amp12_1}
\end{figure}

Most AGB stars vary to some extent, due to radial oscillations in their 
cool, convective mantles. On the thermal-pulsing (TP) upper slopes of the 
AGB, they undergo large amplitude variations on timescales of months 
to years in a (semi-)regular fashion, and are therefore referred 
to as Long Period Variables (LPVs) (e.g., \citet{Ita04, Whi03}).
The 19 month 
time coverage in our data gives enough baseline to detect variations 
given the expected LPVs period range for the majority of 
TP-AGB stars of 100 to 700 days. Indeed, a number of
variable stars were cross-correlated and identified in the
\citet{Har06} catalog where they classify the sources to be
long-period variables based on $r'$, $g- r'$ and $i'$, $r' - i'$
colors. The amplitude range, measured by the standard deviation 
about the mean magnitude, for variable sources in our 
survey at 3.6~$\mu$m is $0.1 \la {\rm amplitude} \la 0.8$ 
(Figure~\ref{fig:amp12_1}).  This range is within
the limits found by \citet{Rej03} at 2.2~$\mu$m of $0.1 \la {\rm amplitude}
\la 2.0$ for Miras and semi-regulars and comparable to the peak-to-peak 
flux variation of 1.4 mag at 10.10~$\mu$m found by \citet{van98}. 
The carbon variable stars show
greater amplitudes than their blue, O-rich counterparts
(Fig.~\ref{fig:amp12_1}).  The oxygen-rich AGB variables are most
likely O-rich Miras which typically are 2--3 visual magnitudes
fainter than other LPVs that are not oxygen-rich. 
This is in good agreement with the
brightness distribution of O-rich and carbon variable stars in
Figure~\ref{fig:cmd12}.
The most extreme cases of TP-AGB stars 
vary on timescales longer than 700 days when they become 
dust-enshrouded and often exhibit OH masers as is the case of oxygen-rich 
objects. These dust-enshrouded objects have a strong mid-IR excess 
and can be preferentially detected over IR-fainter, less evolved AGB 
stars. Our 19 months coverage is sufficient to detect these 
variables as well. Although their amplitudes may have been 
underestimated because the time coverage may have been only half a 
pulsation cycle, these stars exhibit the largest amplitudes of 
all in the IR and will be identified as variable in our catalog 
even though we have sampled only half of their pulsation cycle. 

\begin{figure}
\plotone{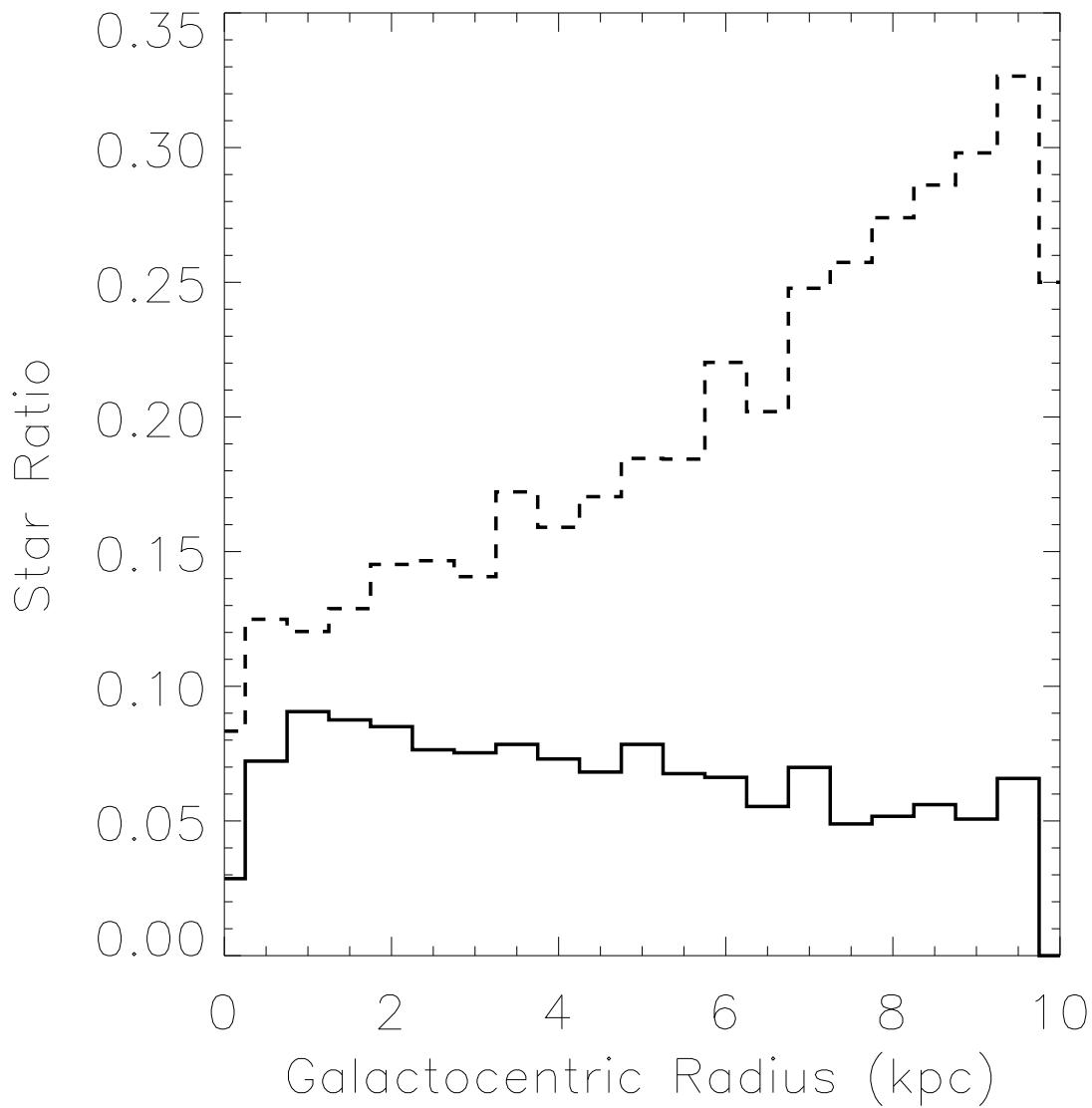}
\caption{Star types versus galacto-centric radius, assuming all stars
lie in the plane of M33.  The solid line shows the ratio of variable
to non-variable stars. The foreground contribution has not been
subtracted, and this most likely accounts for the apparent decreasing
ratio of variables to non-variables at greater distances from the
center of M33. The dashed line shows the ratio of carbon stars to
total number of stars.  The region inside 0.4~kpc is not included
because the surface brightness was too high to detect stars reliably.}
\label{fig:radial_profile}
\end{figure}

The radial profile of variable and non-variable stars 
(Figure~\ref{fig:radial_profile}) shows the ratio of variable stars to 
non-variable stars highest in the inner regions of the galaxy and 
decreasing with radius. This is most likely an effect from the field 
contamination stars. Well outside of the spiral arms ($R > 6$~kpc), the 
percentage of variable stars is roughly constant, indicative of an 
intermediate age AGB stellar population ($\simeq$1--5~Gyr old) that has 
been dynamically mixed in the galaxy.  The number carbon stars detected
increases with galactocentric radius (Figure~\ref{fig:radial_profile}).

\begin{figure}
\plotone{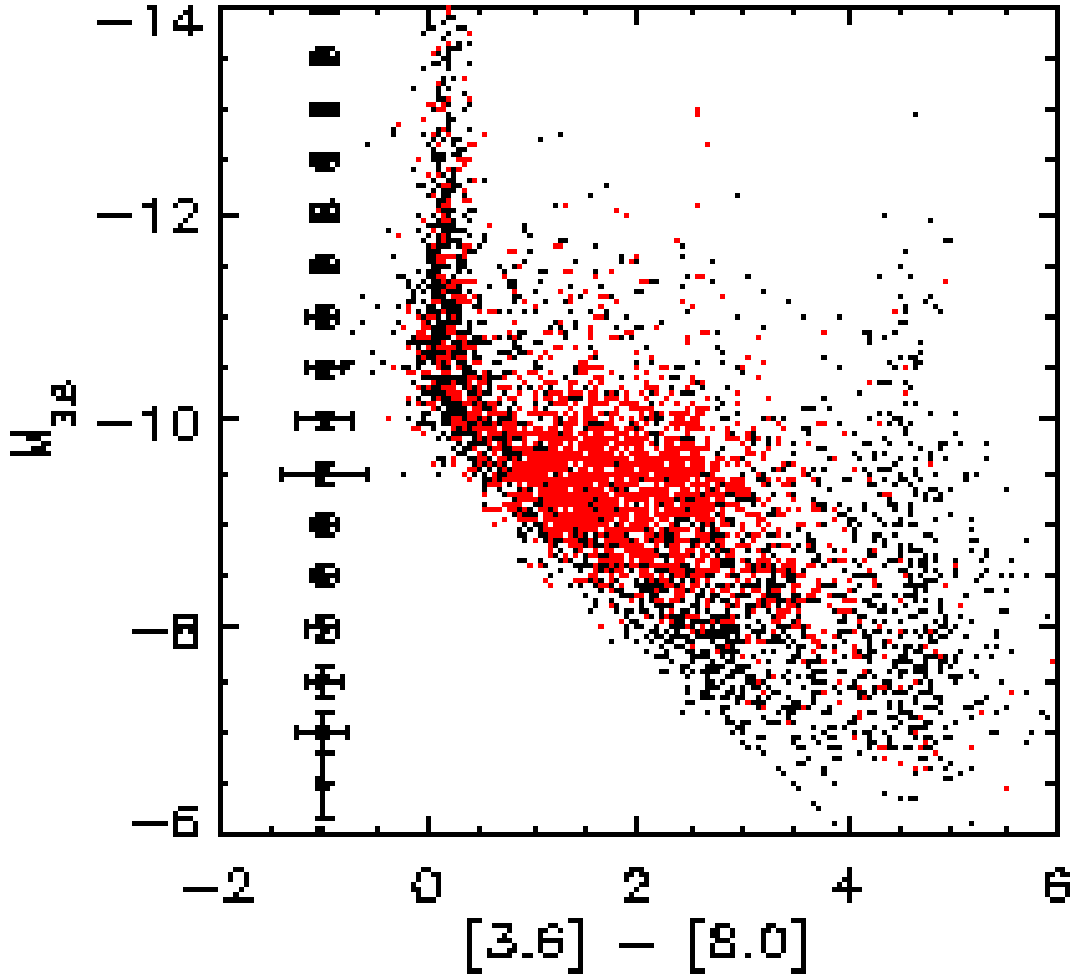}
\caption{Color-magnitude diagram showing point sources in the catalog
at IRAC 3.6 and 8.0~$\mu$m. Non-variable sources are plotted in
black, variable in red. The diagonal line shows the completeness
limit of the survey.  The error bars show representative
uncertainties of a single epoch measurement; they do not vary
significantly with color. The error bars are larger at $M_{3.6}
\approx -9.5$ and $\approx -7$ because many of the 8.0 $\mu$m
magnitudes are near the completeness limit and therefore have larger
uncertainties. This figure contains only 2,689 stars with measurements
in all five epochs at 3.6 or 8.0~$\mu$m.}
\label{fig:cmd14}
\end{figure}

\begin{figure}
\plotone{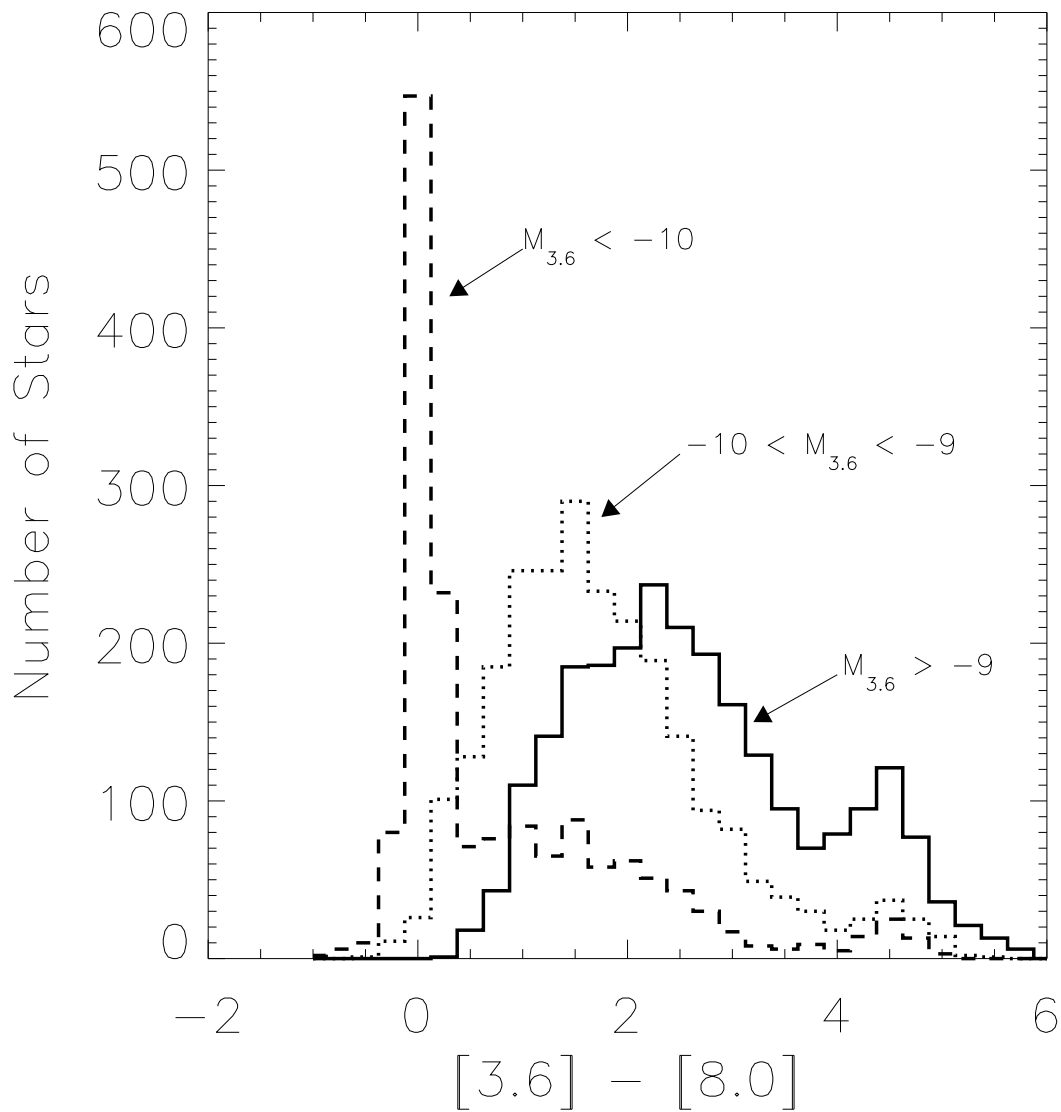}
\caption{Histogram of $[3.6] - [8.0]$ colors for all stars detected 
at 8.0~$\mu$m. Fully 80\% of these
objects show mid-infrared excesses caused by circumstellar dust shells. 
Stars with $M_{3.6}<
-10$ are shown as a dashed line, $-10 < M_{3.6} < -9$ as a dotted
line, and $M_{3.6} > -9$ as a solid line.  Major peaks in the
histograms indicate AGB stars without dust ($[3.6] - [8.0]\approx0$),
carbon stars with dust ($0.5 \ga[3.6] - [8.0]\ga2.5$, and $M_{3.6}<
-9$), and YSOs ($[3.6] - [8.0]\ga4$). }
\label{fig:color14_histogm}
\end{figure}

\begin{figure}
\plotone{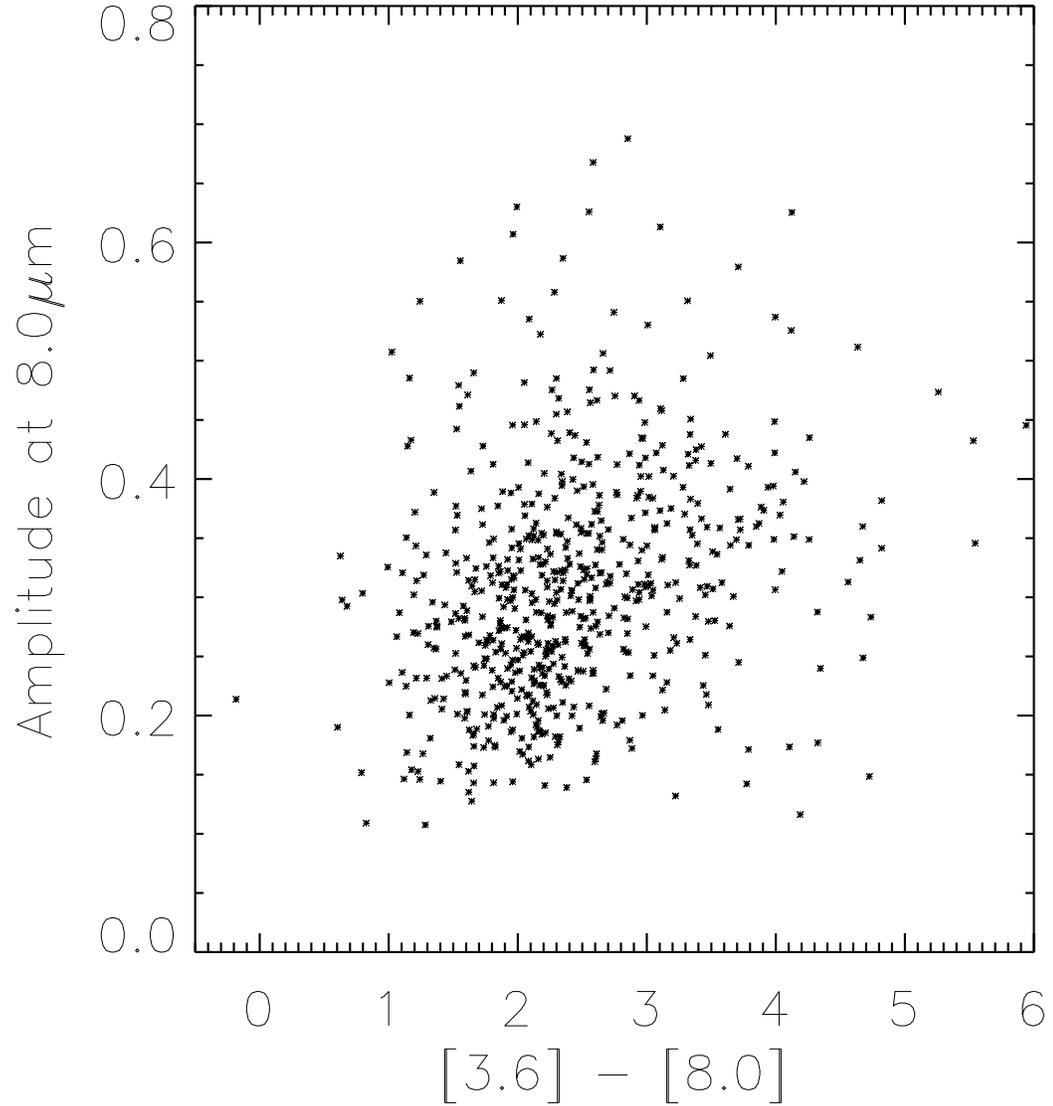}
\caption{Variability amplitude versus color. The amplitude is defined
as the standard deviation of the magnitudes for the five epochs of
observations for each point source. Only 615 stars found to vary at
8.0~\micron\ are plotted.  The amplitudes are somewhat lower than at
the shorter 3.6 $\mu$m wavelength, as expected for dust shell
reradiation.
\label{fig:amp14_4}}
\end{figure}

The carbon stars are bright enough to lie above the IRAC observational 
completeness limit at 8.0~$\mu$m. Their $[3.6] - [8.0]$ colors are shown 
in Figure~\ref{fig:cmd14}. Because both the non-variable and variable 
stars have dust shells giving rise to an IR excess, all but a few of the 
carbon stars have $ 0.5 < [3.6] - [8.0] < 3.5$ . 
The dust shells were created by 
previous or ongoing mass loss that characterizes much of the AGB stage of 
stellar evolution. The mass loss phenomenon is commonly thought to be at 
least partially linked to the radial pulsations and variability of these 
stars. That increased mass loss correlates with increased IR luminosity of 
LPVs \citep{van99} is not surprising. These carbon stars with dust shells 
account for only one part of the tri-modal distribution across 
$[3.6] -[8.0]$ colors of point sources (see 
Figure~\ref{fig:color14_histogm}). Making up the second group are the 
brightest sources at 3.6~$\mu$m in our survey at $[3.6] -[8.0] \approx 0$.  
This population is a mix of field contamination stars and M33 AGB stars 
\citep{Bol07}. The red tail of this bright 
3.6~$\mu$m population is spread over a $[3.6] - [4.5]$ color range of 
$-$0.8 to $+$1.7 indicating it is a mix of oxygen and carbon stars. 
AGB stars with near-zero $[3.6] - [8.0]$ color are the ones 
without significant circumstellar dust shells.  Given the sensitivity of 
the 8.0~\micron\ data, dust shells are only detectable for the most 
luminous AGB stars, those with $M_{3.6} \simeq -10$ (our completeness 
limit at 8.0~\micron\ is $M_{8.0}=-10.5$). Circumstellar shells 
are more difficult to detect around brighter stars as higher mass-loss 
rates are needed to create a similar optical depth dust shell around 
the higher luminosity stars than around a star less luminous 
(e.g., \citet{van07}). The 
third peak in the distribution has $[3.6] - [8.0]> 3.5$.  Objects with 
these colors are mostly young stellar objects (YSOs) according to 
\citet{Bol07}. These objects correlate well with the spiral pattern in M33
and are all located within a 5.7 kpc radius from the 
galaxy's center (Figure \ref{fig:3.6}). 
Only $\sim$15\% of variable stars have $[3.6] - [8.0]$ 
colors redder than 3.5. These stars are most likely LPVs with thick dust 
shells. The amplitudes at 8.0~$\mu$m of the variable stars detected at the 
longer wavelength are shown in Figure~\ref{fig:amp14_4}. The properties of 
the classes are summarized in Table~\ref{tab:prop_var}. 

\section{Conclusions \label{conclusions} }

\textit{Spitzer} IRAC 3.6, 4.5, and 8.0~$\mu$m images give a new look
at the evolved stellar population in M33.  Only stars above the TRGB
are detected in current data, but both oxygen-rich and carbon-rich
AGB stars can be seen. The 2923 variable stars found are most
likely long-period variables with amplitudes as small as 0.1~mag and
as large as 0.8, 1.0, and 0.7~mag at 3.6, 4.5, or 8.0~$\mu$m,
respectively.  Over a third of carbon stars are variable, and those
detected at 8.0~$\mu$m show evidence of dust shells. 
Relatively few of the brightest AGB stars detected at 8.0~$\mu$m showed
evidence of circumstellar dust. 

\acknowledgments
This work is based on observations made with the 
{\it Spitzer Space Telescope,} which is operated by the 
Jet Propulsion Laboratory, 
California Institute of Technology under a contract with NASA. Support for 
this work was provided by NASA through an award issued by JPL/Caltech.  
Support was also provided by NASA through contracts 1256406 and 1215746 
issued by JPL/Caltech to the University of Minnesota. KBWMc also 
acknowledges partial support from the University of Minnesota NASA Space 
Grant Consortium. The authors wish to thank the anonymous
referee whose suggestions have helped improved this manuscript.

{\it Facilities:} \facility{Spitzer (IRAC)}


\clearpage


\input{tab1}

\input{tab2}

\input{tab3}

\input{tab4}

\input{tab5}

\clearpage

\clearpage

\clearpage

\clearpage

\clearpage

\clearpage

\clearpage

\clearpage

\clearpage

\clearpage

\clearpage

\clearpage

\clearpage

\clearpage

\clearpage

\clearpage

\clearpage

\clearpage

\clearpage

\clearpage

\end{document}

%% file: tab1.tex
\clearpage
\begin{deluxetable}{ccccc}
\tablewidth{0pt}
\tablecaption{Journal of Observations \label{tab:obs_summary}}
\tablecolumns{5}
\tablehead{
\colhead{Obs.} & \colhead{Outer AOR} & \colhead{Outer} & \colhead{Inner AOR}
&\colhead{Inner}\\
\colhead{Date} & \colhead{UT Time} & \colhead{AOR No.} & \colhead{UT Time}
& \colhead{AOR No.}}
\startdata
2004/01/09 & 16:01 & 3636224 & 17:17 & 3636480\\
2004/07/22 & 21:43 & 3638016 & 22:56 & 3637760\\
2004/08/16 & 03:02 & 3639040 & 04:15 & 3638784\\
2005/01/21 & 15:14 & 3640576 & 16:24 & 3640320\\
2005/08/25 & 03:00 & 13474048 & 04:13 & 13474304\\
2006/01/09 & 23:09 & 3649792 & 01:12 & 3650048\\
\enddata
\tablecomments{Inner and Outer AORs refer to the two regions M33 was divided
into for observation. The final observations were not used in this study.}
\end{deluxetable}

%% file: tab2.tex
\clearpage
\begin{deluxetable}{c c c c c c c c c c c c c c c c c}
\tablewidth{0pt}
\rotate
\tabletypesize{\scriptsize}
\tablecaption{Catalog of Non-variable Point Sources for M33
\label{tab:catalog1}}
\setlength{\tabcolsep}{0.01in}
\tablehead{
\colhead{SSTM3307} &
\colhead{R.A.} &
\colhead{Decl.} &
\colhead{M$_{3.6}$:1} &
\colhead{$\sigma_{3.6}$:1} &
\colhead{M$_{3.6}$:2} &
\colhead{$\sigma_{3.6}$:2} &
\colhead{...} &
\colhead{M$_{8.0}$:5} &
\colhead{$\sigma_{8.0}$: 5} &
\colhead{$\bar{M}_{3.6}$} &
\colhead{$\bar{M}_{4.5}$} &
\colhead{$\bar{M}_{8.0}$} &
\colhead{[3.6]$-$[4.5]} &
\colhead{[3.6]$-$[8.0]}
}
\startdata
J013151.74+302545.2 & 22.965624 & 30.429226 & 16.075 & 0.042 & 16.049 & 0.061 
& ... & NaN & NaN & 16.105 & 16.038 & NaN & -0.019 & NaN \\
J013152.64+302626.5 & 22.969362 & 30.440706 & 16.092 & 0.087 & 15.999 & 0.057 
&  ... &NaN & NaN & 16.089 & 16.041 & NaN & 0.000 & NaN \\
J013154.34+302628.1 & 22.976429 & 30.441154 & 17.920 & 0.171 & 17.571 & 0.115 
& ... & NaN & NaN & 17.496 & 17.108 & NaN & 0.769 & NaN \\
J013154.76+302559.2 & 22.978176 & 30.433132 & 16.981 & 0.073 & 17.049 & 0.064 
& ... & NaN & NaN & 16.951 & 16.971 & NaN & -0.170 & NaN \\
J013156.45+302312.7 & 22.985233 & 30.386887 & 15.170 & 0.047 & 15.182 & 0.055 
& ... & NaN & NaN & 15.175 & 15.260 & NaN & -0.026 & NaN \\
J013156.49+302304.4 & 22.985384 & 30.384579 & 16.883 & 0.088 & 16.770 & 0.078 
& ... & NaN & NaN & 16.808 & 16.877 & NaN & -0.255 & NaN \\
J013156.50+302335.0 & 22.985420 & 30.393066 & 15.823 & 0.053 & NaN & NaN & ... 
& NaN & NaN & 15.836 & 15.857 & NaN & -0.075 & NaN \\
J013156.71+302623.0 & 22.986320 & 30.439747 & 16.240 & 0.076 & 16.191 & 0.039 
& ... & NaN & NaN & 16.195 & 16.180 & NaN & 0.087 & NaN \\
J013156.80+302348.0 & 22.986694 & 30.396683 & 14.623 & 0.038 & 14.653 & 0.044 
& ... & NaN & NaN & 14.648 & 14.713 & 14.347 & -0.100 & 0.276 \\
J013156.90+302300.9 & 22.987101 & 30.383591 & 17.089 & 0.088 & 16.883 & 0.068 
& ... & NaN & NaN & 16.915 & 17.023 & NaN & 0.213 & NaN \\
\enddata
\tablecomments{The complete version of this table is in the
electronic edition of the Journal. The printed version contains only
a sample. Cols. (2) and (3): R.A. and Decl. in J2000 coordinates,
cols. (4) through (19) (only columns 4, 5, 6, 7, 18, and 19 shown here):
Absolute magnitude and associated photometric
uncertainty given by DAOphot/ALLSTAR at 3.6, 4.5, and 8.0 $\mu$m for each
of the five observational epochs. Col. (20)--(22): Mean magnitude at
each wavelength. Col (23) and (24): IRAC colors.}
\end{deluxetable}

%% file: tab3.tex
\clearpage
\begin{deluxetable}{c c c c c c c c c c c c c c c c c c}
\tablewidth{0pt}
\rotate
\tablecaption{Catalog of Variable Point Sources for M33
\label{tab:catalog2}}
\tabletypesize{\scriptsize}
\setlength{\tabcolsep}{0.01in}
\tablehead{
\colhead{...} &
\colhead{R.A.} &
\colhead{Decl.} &
\colhead{M$_{3.6}$:1} &
\colhead{$\sigma_{3.6}$:1} &
\colhead{M$_{3.6}$:2} &
\colhead{$\sigma_{3.6}$:2} &
\colhead{...} &
\colhead{M$_{8.0}$:5} &
\colhead{$\sigma_{8.0}$: 5} &
\colhead{$\bar{M}_{3.6}$} &
\colhead{$\bar{M}_{4.5}$} &
\colhead{$\bar{M}_{8.0}$} &
\colhead{[3.6]$-$[4.5]} &
\colhead{[3.6]$-$[8.0]} &
\colhead{Amp$_{3.6}$} &
\colhead{Amp$_{4.5}$} &
\colhead{Amp$_{8.0}$}
}

\startdata
.. & 23.007948 & 30.434259 & 15.351 & 0.052 & 15.398 & 0.054 & ... & NaN 
& NaN & 15.545 & 15.413 & NaN & 0.138 & NaN & 0.269 & 0.155 & NaN \\
.. & 23.016241 & 30.418886 & 16.743 & 0.059 & 16.746 & 0.058 & ... & NaN 
& NaN & 16.741 & 16.553 & NaN & 0.142 & NaN & NaN & 0.232 & NaN \\
.. & 23.020172 & 30.466431 & 16.114 & 0.058 & 15.580 & 0.053 & ... & NaN 
& NaN & 15.777 & 15.556 & NaN & 0.465 & NaN & 0.234 & 0.147 & NaN \\
.. & 23.025049 & 30.449522 & 15.576 & 0.046 & 15.812 & 0.039 & ... & NaN 
& NaN & 15.710 & 15.688 & 14.498 & -0.179 & 0.920 & 0.185 & NaN & NaN \\
.. & 23.028364 & 30.645405 & 15.456 & 0.084 & 15.906 & 0.049 & ... & NaN 
& NaN & 15.880 & 16.222 & NaN & -0.702 & NaN & 0.269 & NaN & NaN \\
.. & 23.041285 & 30.486017 & 14.036 & 0.051 & 14.344 & 0.043 & ... & NaN 
& NaN & 14.329 & 14.293 & 13.702 & -0.090 & 0.482 & 0.190 & 0.152 & NaN \\
.. & 23.041332 & 30.442278 & 15.323 & 0.064 & 14.929 & 0.055 & ... & 12.825 
& 0.043 & 15.199 & 14.478 & 12.882 & 0.716 & 2.467 & 0.348 & 0.406 & 0.238 \\
.. & 23.042356 & 30.694159 & 16.391 & 0.109 & 16.369 & 0.052 & ... & NaN 
& NaN & 16.349 & 16.101 & NaN & -0.034 & NaN & NaN & 0.281 & NaN \\
.. & 23.048187 & 30.781109 & 16.672 & 0.100 & 16.712 & 0.035 & ... & NaN 
& NaN & 16.758 & 16.147 & NaN & 0.382 & NaN & NaN & 0.379 & NaN \\
.. & 23.048820 & 30.349766 & 14.791 & 0.038 & 14.897 & 0.044 & ... & NaN 
& NaN & 14.879 & 14.881 & 14.573 & -0.248 & 0.270 & 0.112 & 0.094 & NaN \\
\enddata

\tablecomments{The complete version of this table is in the
electronic edition of the Journal. The printed version contains only
a sample. Col. (1) Star ID is omitted for space. Cols. (2)--(24) are
the same as in Table~\ref{tab:catalog1}.  Cols. (25)--(27): Amplitude
of variability at each wavelength measured by the standard 
deviation about the mean magnitude.}
\end{deluxetable}

%% file: tab4.tex
\clearpage
\begin{deluxetable}{lcccc}
\tablewidth{0pt}
\tablecaption{Number of Stars Detected in M33\label{tab:num_var}}
\tablecolumns{5}
\tablehead{
\colhead{Star Type} &
\colhead{3.6 $\mu$m} &
\colhead{4.5 $\mu$m}  &
\colhead{8.0 $\mu$m} &
\colhead{Total}}
\startdata
All Stars & 37,699 & 37,773 & 5,537 & 39,639 \\
M33 Stars & 29,633  & 29,795  & 4,739 & 32,646 \\
Variable Stars & 1,802 & 2,181 & 615 & 3,047 \\
\enddata
\tablecomments{M33 stars calculated by subtracting the number of
foreground stars from all stars in the catalog. This is an over-
subtraction as the reference fields used to analyze the field
contamination still contain M33 stars. Numbers refer to the
area outside the high surface 
brightness regions ($\simeq0.57$~deg$^2$.)}
\end{deluxetable}

%% file: tab5.tex
\clearpage
\begin{deluxetable}{lcccc}
\tablewidth{0pt}
\tablecaption{Properties of Stars Classes \label{tab:prop_var}}
\tablecolumns{5}
\tablehead{
\colhead{Class of Star} && \colhead{[3.6] - [4.5]} && \colhead{[3.6] - [8.0]}
}
\startdata
Oxygen-rich AGB Star &&  $<$0.2 && $<0.5$\\
Carbon Star && $>$0.2 && $>0.5$ \\
Dust Shell && $\ldots$ && $>$0.5 \\
No Dust && $\ldots$    && $<0.5$ \\
YSO && $\ldots$  && $>$3.5 \\
\enddata
\end{deluxetable}